\documentclass[aip,graphicx,amsmath,nofootinbib]{revtex4-1}

\usepackage{graphicx}
\usepackage{subcaption}
\usepackage{url}

\renewcommand*{\thefootnote}{\fnsymbol{footnote}}

\draft 

\begin{document}

\title{The impact of non-local parallel electron transport on plasma-impurity reaction rates in tokamak scrape-off layer plasmas} 



\author{Dominic Power}
\email[]{power8@llnl.gov}
\affiliation{Blackett Lab., Plasma Physics Group, Imperial College London, London, United Kingdom of Great Britain and Northern Ireland}
\affiliation{Lawrence Livermore National Laboratory, Livermore, California 94550, USA}
\author{Stefan Mijin}
\affiliation{UKAEA, Culham Campus, Oxon, United Kingdom of Great Britain and Northern Ireland}
\author{Kevin Verhaegh}
\affiliation{UKAEA, Culham Campus, Oxon, United Kingdom of Great Britain and Northern Ireland}
\author{Fulvio Militello}
\affiliation{UKAEA, Culham Campus, Oxon, United Kingdom of Great Britain and Northern Ireland}
\author{Robert J. Kingham}
\affiliation{Blackett Lab., Plasma Physics Group, Imperial College London, London, United Kingdom of Great Britain and Northern Ireland}


\date{\today}

\begin{abstract}
  Plasma-impurity reaction rates are a crucial part of modelling tokamak scrape-off layer (SOL) plasmas. To avoid calculating the full set of rates for the large number of important processes involved, a set of effective rates are typically derived which assume Maxwellian electrons. However, non-local parallel electron transport may result in non-Maxwellian electrons, particularly close to divertor targets. Here, the validity of using Maxwellian-averaged rates in this context is investigated by computing the full set of rate equations for a fixed plasma background from kinetic and fluid SOL simulations. We consider the effect of the electron distribution as well as the impact of the electron transport model on plasma profiles. Results are presented for lithium, beryllium, carbon, nitrogen, neon and argon. 
  It is found that electron distributions with enhanced high-energy tails can result in significant modifications to the ionisation balance and radiative power loss rates from excitation, on the order of 50-75\% for the latter. Fluid electron models with Spitzer-H\"{a}rm or flux-limited Spitzer-H\"{a}rm thermal conductivity, combined with Maxwellian electrons for rate calculations, can increase or decrease this error, depending on the impurity species and plasma conditions. Based on these results, we also discuss some approaches to experimentally observing non-local electron transport in SOL plasmas.
\end{abstract}


\maketitle

\section{Introduction}
\label{sec:intro}




Impurity species, elements which are distinct from the fuel mixture in a fusion plasma, are present in all magnetic fusion experiments. They enter either via physical sputtering from wall materials, e.g. beryllium, carbon or tungsten, or are injected deliberately for their favourable radiative properties, where species such as nitrogen, neon or argon are often used.
These impurities can significantly impact the bulk plasma behaviour in both the scrape-off layer (SOL) and the core, and so their behaviour and interaction with the fuel plasma must be understood. Of particular interest are the plasma-impurity reaction rates, where inelastic collisions between impurity particles (neutral atoms or ions) and electrons, particularly ionisation, excitation and their inverse processes, play an important role in determining the SOL power balance and the impurity charge state distribution as well as its average charge. Therefore, capturing these processes accurately is important in understanding the radiative and transport properties of a given impurity species present. 

A common approach to modelling impurities in SOL plasmas is to take effective rate coefficients from a database, for example ADAS, (\url{https://open.adas.ac.uk/}), which typically assumes Maxwellian electrons, and use these as inputs to a transport model for the fuel and impurity plasma species\cite{Sciortino2021,Johnson2019,Reiter2019,Kotov2008}. However, the electrons may be far from Maxwellian in divertor SOL plasmas with moderate to low upstream collisionality and steep temperature gradients, particularly close to the walls, due to non-local parallel transport\cite{Batishchev1999,Tskhakaya2009,Chankin2018,Mijin2020a,Wigram2020,Power2023}.
In particular, the high-energy tail of the electron distribution may be enhanced close to the divertor targets due to the presence of fast particles from the upstream region. Conditions such as this are envisaged in reactor-class devices\cite{Veselova2021,Rubino2017}, and may exist in present-day tokamak experiments\cite{Chankin2009a}. Furthermore, it is generally observed that Langmuir probe electron temperature inferences (which assume a Maxwellian distribution) are overestimated in detached conditions\cite{Batishchev1997,Verhaegh2017}, which has been attributed to the presence of these high-energy tail electrons\cite{Batishchev1996,Batishchev1997}. For the case of electron-impact reactions such as ionisation and excitation, where there is a minimum energy which the incident electrons must possess for the reaction to occur, a modification to the number of particles in the tail of the electron distribution may significantly alter the rate at which the reaction proceeds. 

Attempts to capture the non-local electron transport in fluid SOL models, for example the use of flux limiters\cite{Fundamenski2005} or reduced kinetic models\cite{Brodrick2017}, can lead to modified plasma profiles. Without also accounting for the effect of the modified electron distributions on impurity reaction rates, additional errors may be introduced. Furthermore, it has been suggested by Coster et al.\cite{Coster2011} that a population of fast electrons from upstream in the plasma close to the walls may facilitate divertor detachment at lower upstream densities. This study will attempt to explore this and other questions by using realistic electron distributions from kinetic SOL simulations. A similar effect has been studied recently by Garland et al.\cite{Garland2020,Garland2022}, where the presence of runaway electrons can significantly alter important quantities such as the impurity ion charge state distribution and radiative cooling rates. There, the dominant effect is due to enhanced scattering cross-sections at relativistic energies. There has also been interest in the solar physics community into the effect of suprathermal electrons on reaction rates, treating the fast electron distribution with a power law\cite{smith_enhancement_2003} or as a kappa distribution\cite{dudik_nonequilibrium_2017}. The purpose of the study presented here is primarily to explore the effect of modified electron velocity distributions due to non-local transport, and resulting change to plasma profiles, on reaction rates in relevant SOL regimes. The distributions used are taken from a kinetic calculation of the parallel electron transport, along with fluid models for the background ions and neutral particles, and so the distributions obtained are expected to be relevant to tokamak experiments. It is also worth noting that these distributions are typically most strongly modified in the range of tens to thousands of eV, which is the same energy range in which many important transitions in fusion-relevant impurities live. For example, the ground state single ionisation energies for neon range from 22 eV (Ne$^0 \rightarrow$ Ne$^{+1}$), to 1362 eV (Ne$^{+9} \rightarrow$ Ne$^{+10}$).





In this paper, we will present results of equilibrium impurity atomic state distributions against (fixed) plasma backgrounds from one-dimensional SOL simulations using a kinetic, fluid, and flux-limited fluid transport model for the electrons. These equilibria were obtained without impurities, i.e.\ only electrons and hydrogenic ions and neutrals were modelled. In Section \ref{sec:SIKE}, a collisional radiative model which has been developed for investigations of this type will be presented. It is included here because it has not yet been documented elsewhere, but readers interested in the results may wish to skip this section. In Section \ref{sec:simulations}, the simulations carried out for this study will be outlined. The results are presented in Section \ref{sec:results}. In Section \ref{sec:discussion}, we discuss some of the consequences of the study. There, we show that a bi-Maxwellian can adequately represent the electron distributions used in this study for the calculation of atomic kinetics. We also discuss the possible consequences of non-Maxwellian electrons in detached SOL plasmas, and propose a method of experimental verification. Finally, we summarise the key conclusions of this study in Section \ref{sec:conclusion}.

\section{Collisional radiative model: SIKE}
\label{sec:SIKE}
For the purposes of this study and similar investigations planned for the future, a simple collisional radiative model (CRM) has been developed by the authors. This is called SIKE (\textbf{S}crape-off layer \textbf{I}mpurities with \textbf{K}inetic \textbf{E}lectrons). It is currently under development, but may be accessed publicly at \url{https://github.com/Plasdom/SIKE}. This CRM is similar in nature to other codes in the literature\cite{Johnson2019,Sciortino2021}, with the key difference that electron energy distribution functions can optionally be provided as input as well as electron temperature and density profiles. The FLYCHK code\cite{Chung2005} does have this capability, but SIKE has been developed with large scale analysis of time-dependent 1D plasma profiles and potential future coupling to kinetic codes in mind. The SIKE atomic data input is also designed to be flexible, so that it can be updated or improved over time. 

This CRM will be briefly outlined in this section: the physical model will be outlined first, followed by a description of the atomic data used, and a summary of useful derived quantities which can be calculated by SIKE and which will be used in the analysis of the results in Section \ref{sec:results}. 

\subsection{SIKE model}

The rate coefficient for a collisional process in which impurity particles in atomic state $k$ are produced due to collisions between electrons and particles in state $j$ is labelled $K_{e,j}^k$. This rate coefficient has units [m$^{3}$s$^{-1}$], and is calculated with\cite{Allais2005}
\begin{equation}
  K_{e,j}^k = \frac{4 \pi}{n_e}\int v^3\sigma_{e,j}^k(v) f_0(v) dv,
  \label{eq:SIKE_rate_coeff}
\end{equation}
where $f_0(v)$ is the isotropic part of the electron velocity distribution, which for Maxwellian electrons is $f_0(v)=\left(\frac{m_e}{2\pi kT_e}\right)^{3/2}e^{-m_ev^2/2kT_e}$. 

For a given impurity species, we wish to solve the density evolution equations for each tracked atomic state, assuming at this stage no transport or external sources/sinks.  For an atomic state $k$, with a given ionisation level and electronic configuration, the density evolves according to\cite{Greenland2001} 
\begin{equation}
  \begin{aligned}
  \frac{dn_k}{dt}=&  n_e \sum_{j} n_j K_{e,j}^k +\sum_j A_j^k n_j-n_k n_e \sum_{j} K_{e,k}^j \\
  & - n_k \sum_j A_k^j + n_e \sum_j r^k_jn_j - n_e n_k \sum_j r^j_k,
  \end{aligned}
  \label{eq:SIKE_dens}
\end{equation}
where $K_{e,j}^k$ includes rates from collisional ionisation, excitation and three-body recombination, $A_j^k$ is the radiative transition rate from state $j$ to $k$ (which includes spontaneous de-excitation and auto-ionisation), and $r_j^k$ is the radiative recombination rate from $j$ to $k$. This system of equations defined by (\ref{eq:SIKE_dens}) for all $k$ is written in matrix form,
\begin{equation}
  \frac{d\vec{n}}{dt} = \mathbf{M} \vec{n},
  \label{eq:SIKE_dens_mat}
\end{equation}
where $\vec{n}$ is the vector containing all $n_k$, and the rate matrix $\mathbf{M}$ is filled according to\cite{Summers2006}
\begin{equation}
\begin{aligned}
  M_{jk}&=n_e K^j_{e,k} + A_k^j + n_e r_k^j, \ \ \ \ \ \ k > j \\
  M_{jk}&=n_e K^j_{e,k}, \ \ \ \ \ \ \ \ \ \ \ \ \ \ \ \ \ \ \ \ \ \ \ k < j \\
  M_{jj}&=-\sum_{k \ne j} M_{kj}.
\end{aligned}
\end{equation}
  
We may solve for equilibrium state densities by either evolving (\ref{eq:SIKE_dens_mat}), or by solving the matrix equation directly for $\frac{d\vec{n}}{dt}=0$. In the latter approach, an additional equation specifying the total impurity density is added to (\ref{eq:SIKE_dens_mat}),
\begin{equation}
  n_{imp}^{tot} = \sum_k n_k,
  \label{eq:total_imp_dens_constraint}
\end{equation}
which simply adds a row of ones to $\mathbf{M}$, and means we can solve via 
\begin{equation}
  \vec{n} = \mathbf{M}^{-1}\vec{A},
  \label{eq:SIKE_direct_solve}
\end{equation}
where 
\begin{equation}
  \vec{A}
   = 
   \begin{bmatrix}
      \frac{d\vec{n}}{dt} \\ n_{imp}^{tot}
  \end{bmatrix}
  =
  \begin{bmatrix}
    0 \\ n_{imp}^{tot}
\end{bmatrix}.
\end{equation}
This is equivalent to finding the null space of $\mathbf{M}$, and then applying the constraint given by (\ref{eq:total_imp_dens_constraint}). This equation can be solved using standard matrix solver routines.

\subsection{Atomic data}


The atomic data used in SIKE has been generated using the Flexible Atomic Code (FAC)\cite{Gu2008} for lithium, beryllium, carbon, nitrogen and neon. For argon, a simpler approach using the screened hydrogenic model\cite{Lee1987,Marchand1990} and data from FLYCHK\cite{Chung2005} was used. The two approaches are described in sections \ref{sec:fac_data} and \ref{sec:shm_data} below. In both cases the plasma is assumed to be optically thin.

The atomic data in SIKE is stored in human-readable json files. As such, it may be straightforwardly modified to improve accuracy for future investigations. A description of the input format for atomic data is included with the code repository. 

For each impurity species considered, atomic data is generated for
\begin{itemize}
  \item energy levels,
  \item radiative de-excitation rates,
  \item auto-ionisation rates (spontaneous emission of an electron from a highly-excited atomic state),
  \item radiative recombination cross-sections,
  \item electron-impact ionisation cross-sections (from which we can obtain three-body recombination cross-sections via detailed balance),
  \item electron-impact excitation cross-sections (from which we can obtain de-excitation cross-sections via detailed balance).
\end{itemize}

\subsubsection{FAC data}
\label{sec:fac_data}

Energy levels and oscillator strengths for an ion with a given number of electrons are computed in FAC by diagonalizing the relativistic Hamiltonian\cite{Gaigalas1997,Gaigalas2002}, and collisional cross-sections are treated in the distorted wave approximation\cite{Bar-Shalom1988}. Radiative recombination cross-sections are calculated from the oscillator strengths via the Milne relation\cite{Oxenius1988}. See Gu\cite{Gu2008} for further details on the theory and numerics of FAC.

The cross-section data is generated on an energy grid made of 16 points spaced geometrically from $\varepsilon^\prime = 3.6 \times 10^{-3}$ eV to 500 eV, where $\varepsilon^\prime$ is the post-collision electron energy. We assume cross-sections fall to zero below the incident electron threshold energy for transition. For $\varepsilon^\prime>500$ eV, or for $\varepsilon^\prime$ greater than 200 times the transition energy in the case of excitation, we use the high-energy asymptotic fits provided by FAC. This is a polynomial fit for ionisation and radiative recombination, and the Bethe formula is used for excitation. With this, we can specify the cross-section data on an arbitrary energy grid on which an electron distribution is defined by appropriate translation to the pre-collision energies and log-linear interpolation. 

Atomic data has been computed with FAC for all ionisation stages of helium, lithium, beryllium, boron, carbon, nitrogen, oxygen and neon. Atomic levels are resolved in $n$ (principal quantum number), $l$ (orbital angular momentum quantum number) and $j$ (total angular momentum quantum number). As an example, for neon there are approx. 2500 levels and 300,000 transitions.

Cross-sections for inverse processes are calculated using the principle of detailed balance. For an inelastic process with transition energy $\varepsilon$, the pre- and post-collision velocities are related via 
\begin{equation*}
    \frac{1}{2}m_e v^{\prime 2} = \frac{1}{2}m_e v^2 + \varepsilon.
\end{equation*}
Therefore, for a given excitation cross-section $\sigma^{j^\prime}_{e,j}$ for transitions from state $j$ to $j^\prime$, the de-excitation cross-section is 
\begin{equation}
  \sigma^{j}_{e,j^\prime}(v^{\prime}) = \frac{g_j}{g_{j^{\prime}}} \frac{v^2}{v^{\prime 2}} \sigma^{j^\prime}_{e,j}(v),
  \label{eq:SIKE_sigma_deex}
\end{equation}
where $g_j$ is the statistical weight of state $j$. 
Similarly, three-body recombination cross-sections are calculated from a given ionisation cross-section $\sigma^{j^\prime}_{e,j}$, 
\begin{equation}
  \sigma^{j}_{e,j^\prime}(v^{\prime}) = \frac{1}{2} n_e \lambda^3 \frac{g_j}{g_{j^{\prime}}} \frac{v^2}{v^{\prime 2}} \sigma^{j^\prime}_{e,j}(v),
  \label{eq:SIKE_sigma_3br}
\end{equation}
where $\lambda=\sqrt{\frac{h^2}{2\pi m_ekT_e}}$ is the de Broglie wavelength of an electron at $T_e$. 

\subsubsection{Screened hydrogenic model}
\label{sec:shm_data}

For argon, due to the computational effort required for the full calculation with FAC for large species, a simpler method using the screened hydrogenic model was used. The approach is similar to that used in the FLYCHK code\cite{Chung2005}, and energy levels and spontaneous emission rates are taken from FLYCHK. 

Schematic atomic states are used, distinguished by their principal quantum number $n$. Ionisation cross-sections are calculated with the formula of Burgess and Chidichimo\cite{BurgessChidichimo1983}. Collisional excitation cross-sections for allowed transitions are derived using the van Regemorter formula\cite{Regemorter1962}. In both cases the principle of detailed balance is used to obtain cross-sections for the inverse process (three-body recombination and collisional de-excitation respectively). For radiative recombination, we use the inverse cross-section of the photoionisation process, where Kramers' formula\cite{Kramers1923} is used.  

\subsubsection{Benchmarking}

\begin{figure}[h]
  \centering
  \begin{subfigure}[h]{0.49\textwidth}
      \centering 
      \includegraphics[width=\textwidth]{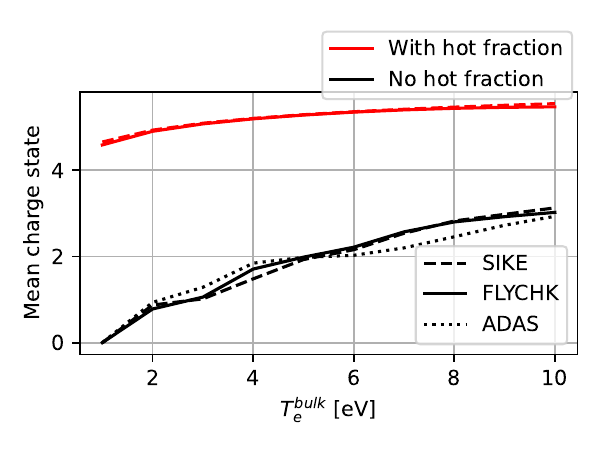}
      \caption{Neon (atomic number $Z = 10$)}
      \label{fig:Ne_hot_frac_FLYCHK_comparison}
  \end{subfigure}
  \hfill
  \begin{subfigure}[h]{0.49\textwidth}
      \centering 
      \includegraphics[width=\textwidth]{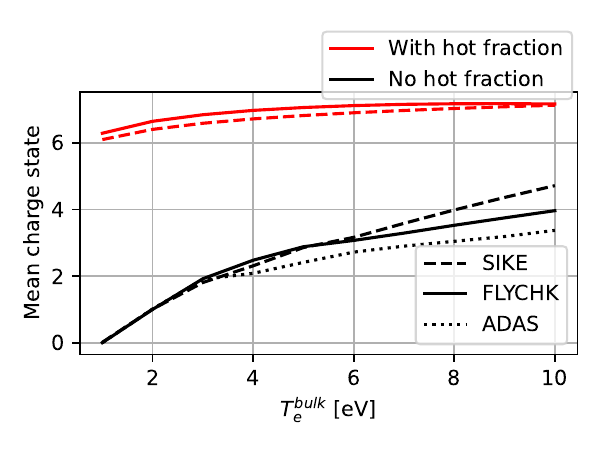}
      \caption{Argon (atomic number $Z = 18$)}
      \label{fig:Ar_hot_frac_FLYCHK_comparison}
  \end{subfigure}
  \caption{Mean charge state at equilibrium, calculated with SIKE, FLYCHK, and ADAS for carbon and neon in the presence of electrons at a range of temperatures with and without a hot fraction (1\%) at 100 eV.}
  \label{fig:hot_frac_FLYCHK_comparison}
\end{figure}


The data used in SIKE is fully self-consistent, meaning both directions of reversible transitions are included, cross-section thresholds are consistent with transition energies, etc. However, energy levels are typically only accurate to within a few eV in FAC. FAC does offer routines to correct for energy levels, but this has not yet been applied to the SIKE data. 

In Figure \ref{fig:hot_frac_FLYCHK_comparison}, we compare the mean charge of neon and argon ions in the presence of background electrons at a density $n_e=10^{20}$m$^{-3}$ for SIKE and FLYCHK. We plot results with and without a fraction 1\% of hot electrons at $T_e^{hot}=100$ eV, where the bulk electron temperature ranges from $T_e^{bulk}=1$ to 10 eV. Results calculated with ADAS are also shown for comparison in the cases without the hot fraction\footnote{For a bi-maxwellian electron distribution, $f=f_{bulk} + f_{hot}$, the effective ionisation and recombination coefficients provided by ADAS are non-linear in $f$.}. There is reasonable agreement between the models. Differences are expected to be due to differences in the atomic data used as input to each model, and this may be improved in future versions of SIKE. 

\subsection{SIKE outputs and useful derived quantities}

Here we will summarise some useful derived quantities from the SIKE model, which can be calculated by SIKE and will be used for the analysis presented later in this paper.

In many cases we can avoid solving for the densities of all states by using generalized collisional radiative theory\cite{Bates1962,Summers2006,Greenland2001}. 
Following Greenland and Reiter\cite{Greenland1998}, we divide the impurity atomic states into $P$ states, which are evolved, and $Q$ states, with zero time derivatives, allowing us to re-write (\ref{eq:SIKE_dens_mat}),
\begin{equation}
  \frac{d}{dt}
  \begin{bmatrix}
      \vec{n}_P \\ \vec{n}_Q
  \end{bmatrix}
   = 
   \begin{bmatrix}
      \mathbf{M}_{PP} & \mathbf{M}_{PQ} \\ \mathbf{M}_{QP} & \mathbf{M}_{QQ}
  \end{bmatrix}
  \begin{bmatrix}
      \vec{n}_P \\ \vec{n}_Q
  \end{bmatrix}
  = 
  \begin{bmatrix}
      \frac{d \vec{n}_P}{d t} \\ 0
  \end{bmatrix}
\end{equation}
and so
\begin{equation}
  \frac{d\vec{n}_P}{dt} = \mathbf{M}_{\text{eff}} \vec{n}_P,
  \label{eq:SIKE_dens_cr}
\end{equation}
where 
\begin{equation}
  \mathbf{M}_{\text{eff}} = \mathbf{M}_{PP} - \mathbf{M}_{PQ} \mathbf{M}_{QQ}^{-1}\mathbf{M}_{QP}
\end{equation}
is the effective rate matrix.

A useful choice of P states is the ground states of each ionisation stage of the impurity species, in which case the lower off-diagonal elements of $\mathbf{M}_{\text{eff}}$ are the effective ionisation coefficients, and the upper off-diagonal elements are the effective recombination coefficients. Defined explicitly in terms of the elements of $\mathbf{M}$, the effective ionisation coefficient is
\begin{equation}
  K_{e\!f\!f}^{ion,z} = K^{z^+}_{e,z} - \vec{K}^{z^+}_{e,i^{\prime}} \mathbf{M}_{ii^{\prime}}^{-1} \mathbf{M}_{iz},
  \label{eq:eff_iz_coeff}
\end{equation}
which describes ionisation from a given ground state $z$ to a ground state of a higher ionisation stage $z^+$. Here, $i$ and $i^{\prime}$ are indices of all non-ground states in the same ionisation stage as $z$, and $\vec{K}^{z^+}_{e,i^{\prime}}$ is the array of all possible ionisation coefficients from states $i^{\prime}$ to $z^+$. Similarly, the effective recombination coefficients from a ground state $z$ to that of a lower ionisation stage $z^-$ is 
\begin{equation}
  K_{e\!f\!f}^{rec,z} = r_{z}^{z^-} - \mathbf{M}_{zi^{\prime}} \mathbf{M}_{ii^{\prime}}^{-1} \vec{r_i}^{z^-},
  \label{eq:eff_rec_coeff}
\end{equation}
where $\vec{r_i}^{z^-}$ is the array of all recombination coefficients from $i$ to $z^-$.

If we evolve (\ref{eq:SIKE_dens_cr}) in time, we will arrive at the same equilibrium as solving the larger set of equations in (\ref{eq:SIKE_dens_mat}), where the densities of the $Q$ states can be extracted via 
\begin{equation}
  \vec{n}_Q = -\mathbf{M}_{QQ}^{-1} \mathbf{M}_{QP} \vec{n}_P.
\end{equation}


The density of a given ionisation stage is 
\begin{equation}
  n_z=\sum_{k \in Z} n_k,
\end{equation}
where $Z$ is the set of all $z$-ionised states. 
The mean charge for the impurity species as a whole is then,
\begin{equation}
  \bar{z} = \frac{\sum_z n_z z}{n_{z,tot}},
\end{equation}
where the summation is over the densities of each ionisation stage multiplied by its ionisation $z$ (i.e. levels with electric charge $+ez$). The total impurity density is $n_{z,tot}=\sum_z n_z = \sum_k n_k$. 

Another useful output is the electron energy loss due to inelastic excitation collisions with the impurity species, where we are typically interested in the contribution from each ionisation stage separately. 
At equilibrium, this is
\begin{equation}
  Q_z = \sum_{k,k^\prime \in Z} A_{k}^{k^\prime} n_k \varepsilon_k^{k^\prime},
  \label{eq:Q_z}
\end{equation}
where $\varepsilon_k^{k^\prime}$ is the transition energy from state $k$ to $k^\prime$, and we have a sum over all radiative de-excitation transitions from all $z$-ionised levels (again denoted by the set $Z$). This has units [Wm$^{-3}$]. We may write this equivalently as 
\begin{equation}
  Q_z= n_e \left( \sum_{k>k^\prime} K_{e,k}^{k^\prime} n_k \varepsilon_k^{k^\prime} - \sum_{k<k^\prime} K_{e,k}^{k^\prime} n_k \varepsilon_k^{k^\prime} \right),\ \mathrm{for}\  k,k^\prime \in Z.
  \label{eq:Q_z_2}
\end{equation}
Normalising to the electron and total impurity density gives a line emission coefficient,
\begin{equation}
  L_z = \frac{Q_z}{n_e n_z},
  \label{eq:cooling_curve_per_ion}
\end{equation}
with units [Wm$^{3}$].
It is also useful to aggregate both these quantities for all ionisation stages, such that the total radiated power is 
\begin{equation}
  Q_{z,tot} = \sum_z Q_z,
  \label{eq:Q_z_tot}
\end{equation}
which is simply summed over all ionisation stages, and the average line emission per ion is
\begin{equation}
  \bar{L}_{z} = \frac{\sum L_z n_z}{n_{z,tot}},
  \label{eq:eff_cooling_curve}
\end{equation}
where $n_z=\sum_{k \in Z}n_k$ is the density of particles of ionisation stage $z$. 

There is an additional contribution to radiative losses from recombination, but this is small in all of the plasma conditions studied here and so is neglected in the analysis. Note that this would not necessarily be the case in SOL transport simulations with impurities included, where the additional radiative losses could reduce the fuel plasma tmeprature such that recombination dominates. 

\section{Simulations}
\label{sec:simulations}


For a given plasma background, SIKE allows us to compute the impurity atomic state distribution. The plasma backgrounds used in this study come from equilibrium simulations with the SOL-KiT code, originally developed by Mijin et el.\cite{Mijin2021} and which has recently been upgraded by Power et al.\cite{Power2023,Power2021}. This is a 1D plasma transport code where the electrons may be treated kinetically, by solving the Vlasov-Fokker-Planck-Boltzmann (VFPB) equation, or as a fluid, using a Braginskii-like transport model\cite{Braginskii2004} with Spitzer-H\"{a}rm heat conductivity\cite{Spitzer1953}. The fuel ions and neutral hydrogenic particles are treated with fluid models. All equations are solved along the direction parallel to the magnetic field, labelled the $x$-axis. 

The SOL-KiT simulations used here are a subset of those presented in the study by Power et al.\cite{Power2023}. A 100\% deuterium plasma was simulated. In kinetic mode, SOL-KiT solves the spherical harmonic expansion of the VFPB equation\cite{shkarofsky1966} under the assumption of azimuthal symmetry of the electron distribution about the $v_x$-axis, which is parallel to the magnetic field. Fokker-Planck and inelastic Boltzmann collision operators are implemented. This series of equations is then solved up to a maximum harmonic, $l_{max}$, which for the simulations presented here is $l_{max}=3$. In fluid mode, a Braginskii-like model is used for the electrons, where we solve for the electron density, $n_e$, flow velocity, $u_e$, and temperature, $T_e$.

Quasi-neutrality is enforced so the electron and ion densities are equal, $n_e=n_i$. Braginskii-like equations are solved for the ion velocity, $u_i$, and ion temperature, $T_i$. A quasi-2D, Navier-Stokes like model is used for the neutral deuterium atoms, where the neutral parallel velocity, $u_{n\parallel}$, perpendicular velocity, $u_{n\perp}$, and temperature, $T_n$, are evolved. The electric field is calculated by solving Amp\`{e}re-Maxwell's law, containing only the displacement current. 

In the fluid version of SOL-KiT, the parallel electron conductive heat flux is given by
\begin{equation}
  q_{SH} =  -\kappa \nabla_{\parallel}T_e,
\end{equation}
where $\kappa$ is the Spitzer-H\"{a}rm heat conductivity. An addition to SOL-KiT for the purposes of this study is a flux limiter, which is applied to the parallel electron conductive heat flux to limit it to some fraction $\alpha$ of the free streaming heat flux, $q_{FS}\simeq 0.8 n_ekT_ev_{th,e}$, where $v_{th,e}$ is the electron thermal velocity. The electron heat flux then becomes
\begin{equation}
  q_{FL} = q_{SH} / (1 + |q_{SH}/\alpha q_{FS}|).
\end{equation}
For this study we use $\alpha=0.2$ as a representative value used in SOL simulations\cite{Fundamenski2005} (although note the range of values for $\alpha$ highlighted in that study; first-principles estimates for $\alpha$ are not readily available). 

Symmetry about the upstream midplane, at $x=0$, is assumed. At the sheath boundary, the Bohm criterion is applied to the plasma flow velocity, $u_e=u_i=c_s$, where $c_s$ is the plasma sound speed. In kinetic mode this boundary condition is applied by truncating the backwards-travelling part of the electron distribution at a cut-off velocity which yields ambipolar flux across the sheath. In fluid mode, a classical sheath heat transmission factor is used\cite{Stangeby2001}. 

The SOL-KiT simulations used for this study are for a SOL with connection length (measured from the sheath entrance to the upstream midplane) $L=30.96m$, and total input power $q_{in}=128$ MWm$^{-2}$. This is inserted equally between the electrons and ions, spatially uniformly over the first third of the domain. The line-averaged total (plasma plus neutral) density, $\langle n \rangle$, was varied from $3 \times 10^{19}$ to $14 \times 10^{19}$ m$^{-3}$. This gives a range of upstream electron collisionality parameters, $\nu_{e,u}^* \simeq 10^{16}n_u / T_{e,u}^2$ from 12 to 51 (where $n_u$ is in m$^{-3}$ and $T_{e,u}$ is in eV). The simulations were run for kinetic, fluid and flux-limited fluid electron transport models. These simulation conditions are not machine-specific, but are instead supposed to represent conditions broadly relevant to present and future devices. See Power et al.\cite{Power2023} for further discussion of the relevance of these conditions. 

\begin{table}[h]
  \centering
  \begin{tabular}{ l|l|l } 
    Label & Transport model & $f_0(v)$ used in rate calculations \\ 
    \hline
    \hline
    kinetic & Kinetic & Evolved $f_0$ from SOL-KiT \\ 
    \hline
    kin. profile & Kinetic & Maxwellian \\ 
    \hline
    SH & Fluid with Spitzer-H\"{a}rm conductivity & Maxwellian \\ 
    \hline
    FL & Fluid with flux-limited SH conductivity & Maxwellian 
  \end{tabular}
  \caption{Four methods of treating the electrons in the study presented here.}
  \label{tbl:electron_treatments}
\end{table}

\begin{figure}[h]
  \centering
  \begin{subfigure}[h]{0.49\textwidth}
      \centering 
      \includegraphics[width=\textwidth]{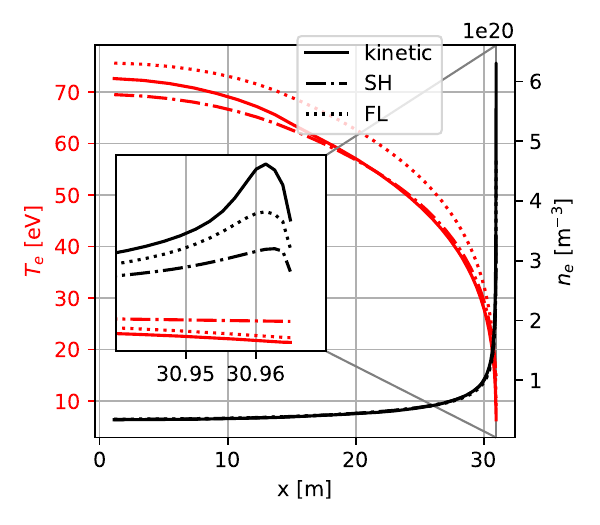}
      \caption{}
      \label{fig:impurity_runs_example_profile}
  \end{subfigure}
  \hfill
  \begin{subfigure}[h]{0.49\textwidth}
      \centering 
      \includegraphics[width=\textwidth]{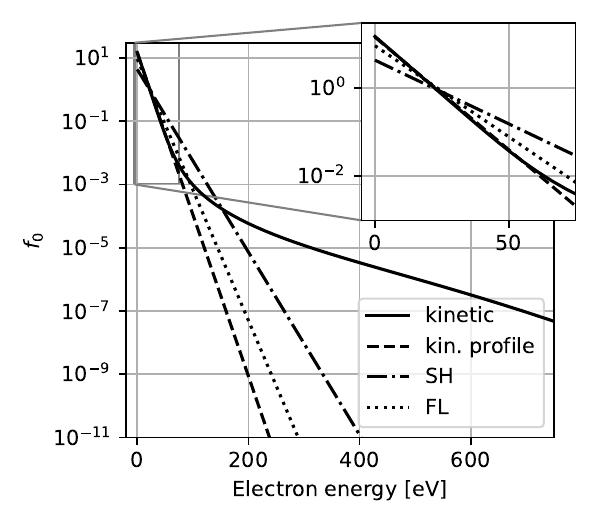}
      \caption{}
      \label{fig:impurity_runs_example_dist}
  \end{subfigure}
  \caption{(a): Example temperature and density profiles of a kinetic and fluid simulation used in this study with $\nu_{e,u}^*=19.8$. Inset is the region close to the target. (b): Isotropic part of the electron distribution close to the target for this example simulation. Shown is the evolved distribution in the kinetic simulation (`kinetic'), a Maxwellian at the local temperature and density in the kinetic simulation (`kin. profile'), and Maxwellians at the temperature and density of the same location from the fluid run with Spitzer-H\"{a}rm conductivity (`SH') and with a flux-limiter (`FL').}
  \label{fig:impurity_runs_example}
\end{figure}

Figure \ref{fig:impurity_runs_example} shows an example plasma profile and electron distribution for one of these simulations at $\langle n \rangle = 5 \times 10^{19}$ m$^{-3}$. Profiles are shown for the kinetic electron transport model (`kinetic'), the fluid model with Spitzer-H\"{a}rm heat conductivity (`SH'), and the flux-limited fluid model (`FL').  Four electron velocity distributions are shown in Figure \ref{fig:impurity_runs_example_dist}: the distribution at a location close to the target from the kinetic run, an equivalent Maxwellian at the same local temperature and density (`kin. profile'), as well as Maxwellians at the same location (but different temperature and density) from the SH fluid profile and the FL fluid profile. These methods of treating the electrons are summarised in Table \ref{tbl:electron_treatments} and will be used as the basis for the forthcoming analysis, and will allow us to observe differences due solely to the departure of the distribution from Maxwellian as well as from the impact of the electron transport model on the plasma profiles.

To compute the impurity atomic state densities, outputs from the SOL-KiT simulations were provided as input to SIKE. For the kinetic SOL-KiT simulations, these outputs were the electron distribution and velocity grid. The impurity state densities were then solved for both this electron distribution and Maxwellian electrons with the same $T_e$ and $n_e$ profiles. For SOL-KiT simulations with fluid electrons (the SH and FL cases), the electron temperature and density profiles were provided as input to SIKE, and the distributions were assumed to be Maxwellian on the same velocity grid as the kinetic runs. The impurity species' treated were lithium, beryllium, carbon, nitrogen and neon. 

The total impurity density was set to 1\% of the local electron density in all runs. Note that this parameter parameter does not affect the results presented here given the modelling assumptions made (fixed background electron density, no impurity transport, optically thin plasma, no effect of impurities on bulk plasma, etc.). Thus we are isolating the impact of kinetic effects and ignoring any impacts on impurity transport, as well as any interactions between these two effects. 

\section{Results}
\label{sec:results}

Although atomic equilibria have been calculated for six impurity species, the results presented here will primarily be for a representative subset of these species. In addition, unless stated otherwise, the plasma profile presented in Figure \ref{fig:impurity_runs_example_profile} will be used as a basis for much of the analysis. 

\subsection{Ionisation balance}

\begin{figure}
  \centering
  \begin{subfigure}{0.49\textwidth}
      \centering 
      \includegraphics[width=\textwidth]{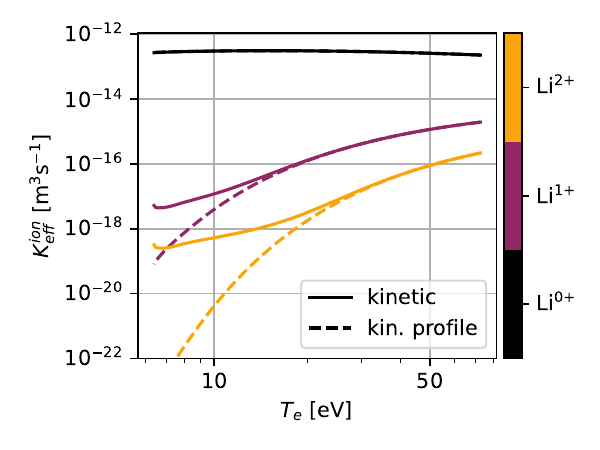}
      \caption{Lithium}
      \label{fig:Li_iz_coeffs}
  \end{subfigure}
  \begin{subfigure}{0.49\textwidth}
      \centering 
      \includegraphics[width=\textwidth]{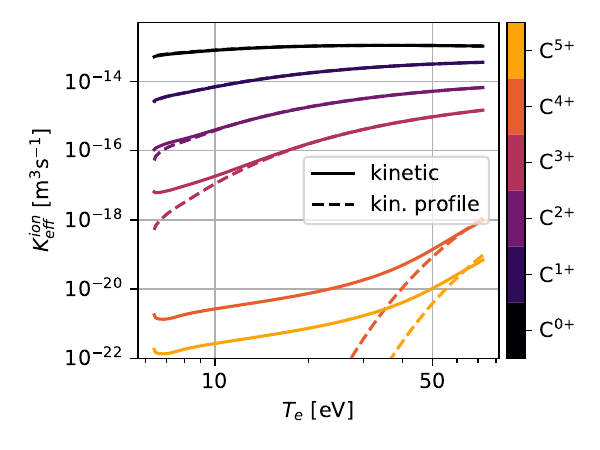}
      \caption{Carbon}
      \label{fig:C_iz_coeffs}
  \end{subfigure}
  \begin{subfigure}{0.49\textwidth}
    \centering 
    \includegraphics[width=\textwidth]{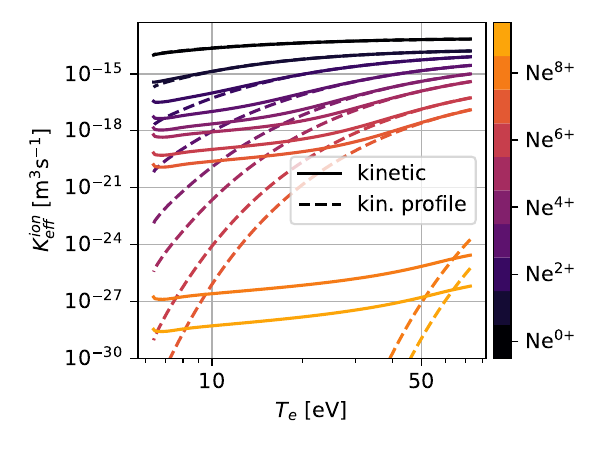}
    \caption{Neon}
    \label{fig:Ne_iz_coeffs}
\end{subfigure}
\begin{subfigure}{0.49\textwidth}
    \centering 
    \includegraphics[width=\textwidth]{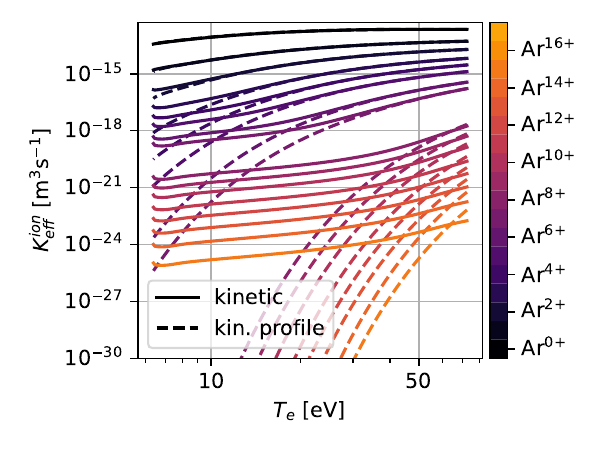}
    \caption{Argon}
    \label{fig:Ar_iz_coeffs}
\end{subfigure}
  \caption{Effective ionisation coefficients for lithium, carbon, neon and argon, where the plasma background is the kinetic run shown in Figure \protect\ref{fig:impurity_runs_example}. Solid lines are for the electron distribution from the SOL-KiT simulation, and dashed lines are for Maxwellian electrons. The $X^{Z\!+}$ labels refer to the particle before ionisation, i.e. the process $X^{Z\!+} \rightarrow X^{(Z\!+\!1)\!+}$.}
  \label{fig:iz_coeffs}
\end{figure}


We start by focussing on the plasma equilibria obtained with the kinetic electron transport model, and investigate the effect of the departure of the electrons from Maxwellian, keeping the density and temperature profiles fixed. 

In Figure \ref{fig:iz_coeffs} we show the effective ionisation coefficients (\ref{eq:eff_iz_coeff}) against $T_e$ for a single plasma profile. The solid lines (`kinetic') are the values of $K_{e\!f\!f}^{ion}$ calculated using the evolved electron distribution from the SOL-KiT simulation, while the dashed lines (`kin. profile') are the values of $K_{e\!f\!f}^{ion}$ for Maxwellian electrons at the same density and temperature. There is clearly strong enhancement to the ionisation rates at low temperatures, particularly for highly-ionised impurity species. This is a result of the strongly enhanced tail of the electron velocity distribution close to the divertor targets in these simulations, where $T_e$ is lowest. Ionisation rates for the neutral and near-neutral species, where low-energy electrons dominate the rate coefficients, show good agreement. For the same reason, the recombination rates for all species at all temperatures are in close agreement.

\begin{figure}
  \centering 
  \includegraphics[width=0.7\textwidth]{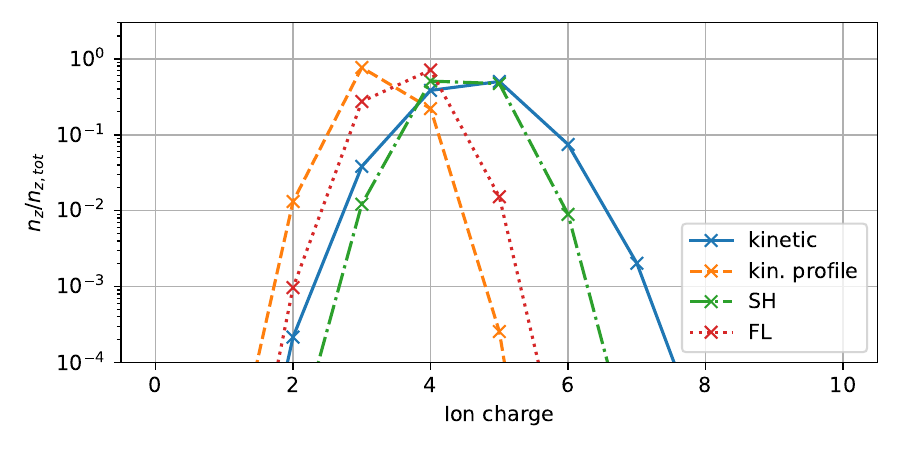}
  \caption{
    Fractional abundance of the ionisation stages of neon at a single location close to the target ($x=30.95$m) for the kinetic transport model and using the electron distribution from SOL-KiT (`kinetic'), the kinetic transport model and Maxwellian electrons (`kin. profile'), the fluid transport model with Maxwellian electrons (`SH') and the flux-limited fluid model with Maxwellian electrons (`FL'); see Table \ref{tbl:electron_treatments} for definitions.
    }
  \label{fig:single_loc_iz_balance_Ne}
\end{figure}

This analysis show how the electron distribution affects the ionisation and recombination rates. However, the choice of transport model used will also affect the plasma profiles obtained. Both of these will influence the ionisation balance of a given impurity species obtained in simulations. In Figure \ref{fig:single_loc_iz_balance_Ne}, the ionisation balance for neon at a single location close to the target ($x=30.95$m, 1cm from the target) is shown for the four ways of treating the electrons shown in Table \ref{tbl:electron_treatments}. The difference between the `kinetic'/`kin. profile' curves is purely due to the electron distribution, while the difference between the `kinetic'/`SH' and `kinetic'/`FL' curves are due to differences in both the distributions and the plasma profiles obtained with these different transport models. It is seen here that a flux-limited fluid model shows larger deviation in mean ionisation with the fully kinetic treatment than a purely fluid model, suggesting a flux limiter may increase errors in ionisation rates used in some scenarios. 

\begin{figure}
  \centering
  \begin{subfigure}{0.8\textwidth}
      \centering 
      \includegraphics[width=\textwidth]{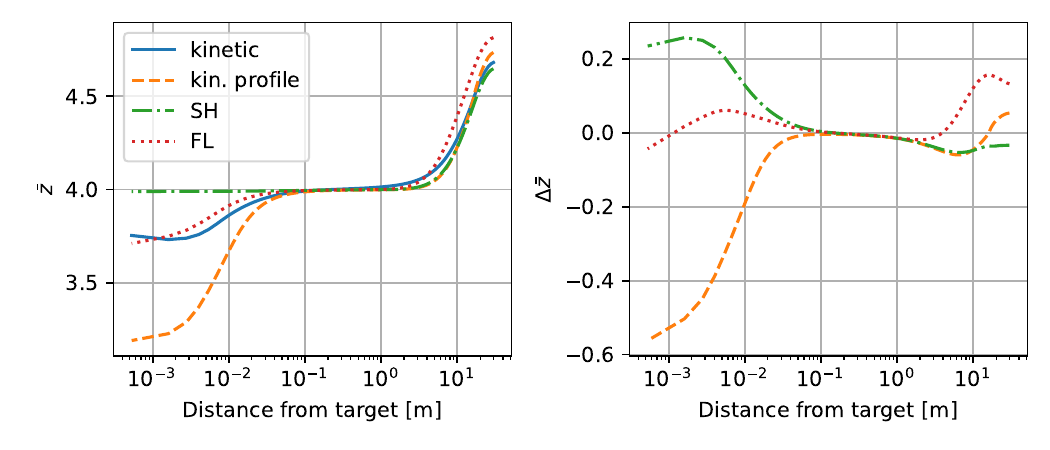}
      \caption{Carbon}
      \label{fig:avg_iz_and_error_C}
  \end{subfigure}
  \hfill
  \begin{subfigure}{0.8\textwidth}
      \centering 
      \includegraphics[width=\textwidth]{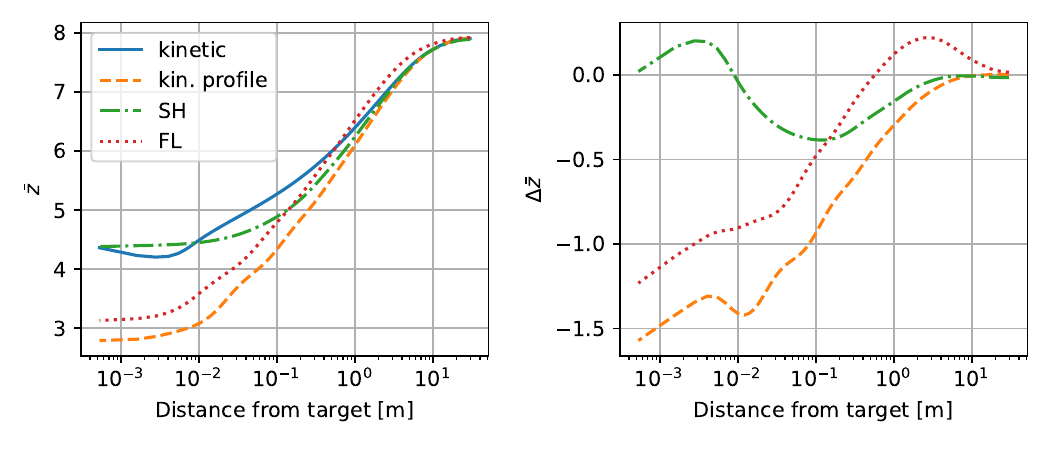}
      \caption{Neon}
      \label{fig:avg_iz_and_error_Ne}
  \end{subfigure}
  \hfill
  \begin{subfigure}{0.8\textwidth}
      \centering 
      \includegraphics[width=\textwidth]{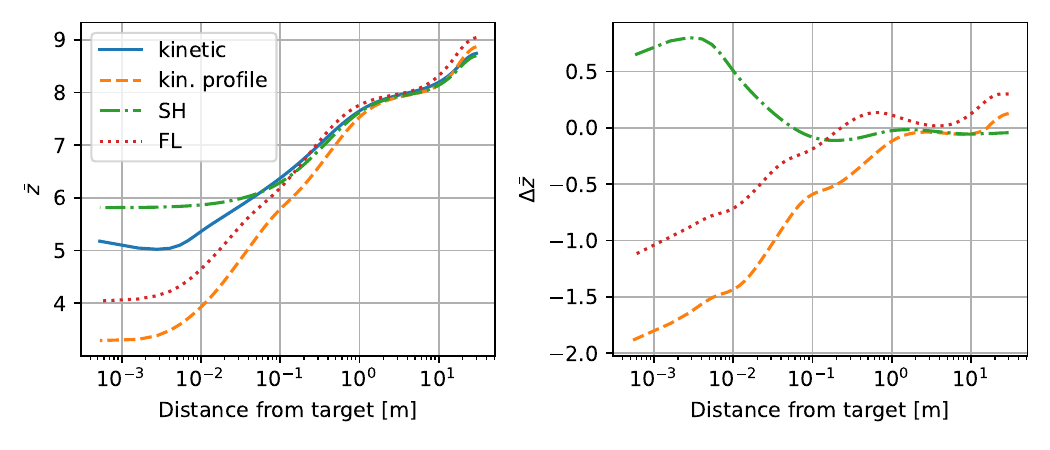}
      \caption{Argon}
      \label{fig:avg_iz_and_error_Ar}
  \end{subfigure}
  \caption{Spatial profiles of the mean charge (left) and error compared to the fully kinetic calculation (right).}
  \label{fig:avg_iz_errors}
\end{figure}

In Figure \ref{fig:avg_iz_errors} we plot the spatial profiles of the mean charge, $\bar{z}$, as well as the deviation in each model compared to the fully kinetic calculation, i.e. $\Delta \bar{z} = \bar{z}^{\cdots} - \bar{z}^{kinetic}$, where the ellipsis is one of `kin. profile', `SH' or `FL'. We have shown results here for carbon, neon and argon, plotted as a function of the distance from the divertor target. We see that the story is different depending on the species considered - for carbon, a flux-limited fluid transport model is closer to the fully kinetic result than the fully fluid model, while the opposite is true in the case of neon and argon. For all impurity species considered, the largest differences arise when the electron transport is treated kinetically, but Maxwellian electrons are used to compute the ionisation balance (the `kin. profile' case).


\subsection{Radiative power loss rate}

\begin{figure}
  \centering
  \begin{subfigure}{0.49\textwidth}
    \centering 
    \includegraphics[width=\textwidth]{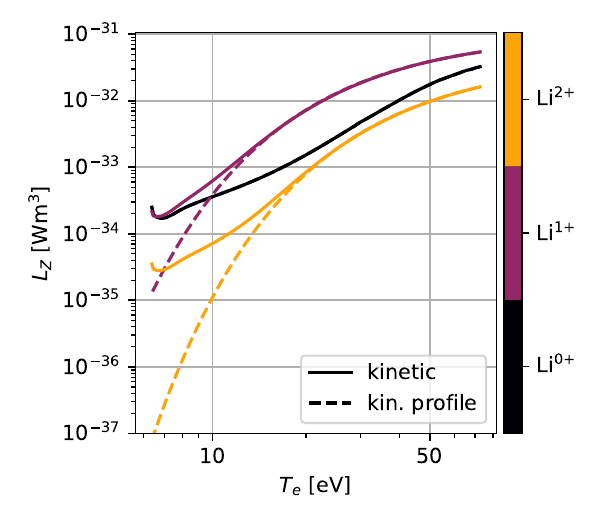}
    \caption{Lithium}
    \label{fig:Li_cooling_curve}
  \end{subfigure}
  \hfill
  \begin{subfigure}{0.49\textwidth}
    \centering 
    \includegraphics[width=\textwidth]{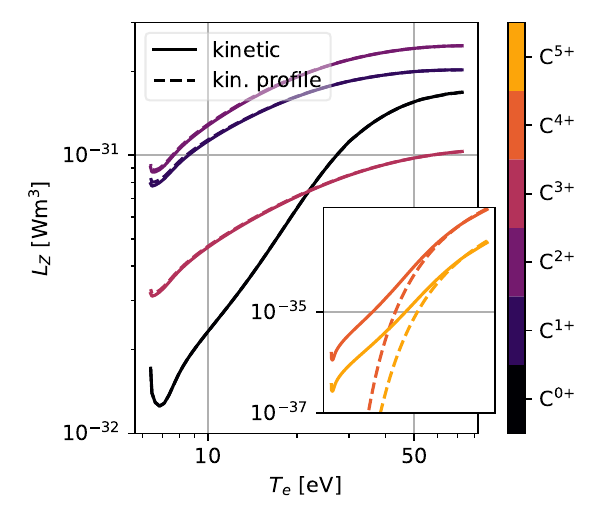}
    \caption{Carbon}
    \label{fig:C_cooling_curve}
  \end{subfigure}
  \centering
  \begin{subfigure}{0.49\textwidth}
    \centering 
    \includegraphics[width=\textwidth]{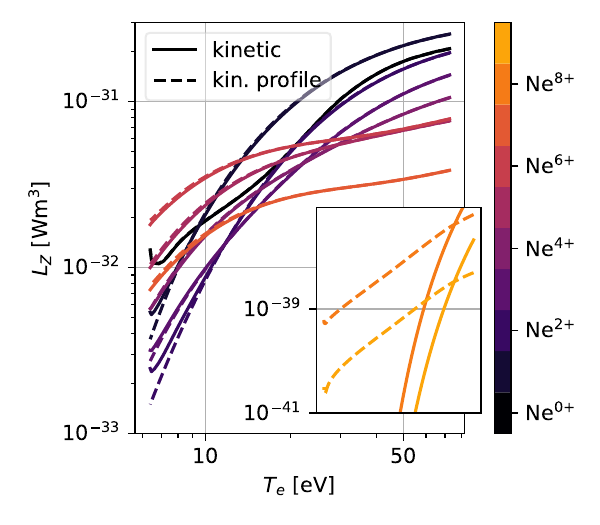}
    \caption{Neon}
    \label{fig:Ne_cooling_curve}
  \end{subfigure}
  \hfill
  \begin{subfigure}{0.49\textwidth}
    \centering 
    \includegraphics[width=\textwidth]{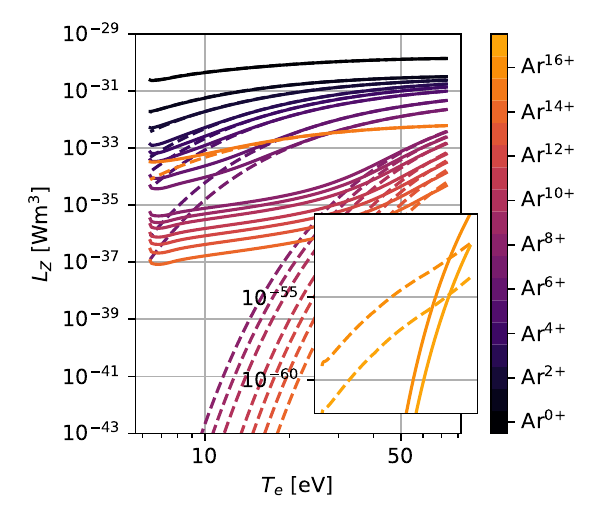}
    \caption{Argon}
    \label{fig:Ar_cooling_curve}
  \end{subfigure}
  \caption{Excitation radiation per ion for lithium, carbon, neon and argon for kinetic and Maxwellian electrons on a plasma profile from the kinetic SOL-KiT run in Figure \protect\ref{fig:impurity_runs_example}.}
  \label{fig:cooling_curves_per_ion}
\end{figure}

In Figure \ref{fig:cooling_curves_per_ion}, we show the excitation radiation per ion, $L_z$ (\ref{eq:cooling_curve_per_ion}), for each ionisation stage of lithium, carbon, neon and argon on this same plasma profile, again investigating the effect of the electron distribution. 
Similarly to the ionisation rates, there are enhancements to the rates of energy loss due to excitation when the evolved electron distribution is used, particularly in the cold regions close to the target. These enhancements are strongest in the hydrogen-like (Li$^{2+}$, Be$^{3+}$, ...) and helium-like (Li$^{1+}$, Be$^{2+}$, ...) ions, most likely due to the fact that atomic energy level spacings are larger in these highly-ionised stages, meaning tail electrons are contributing more to the collisional excitation rates. In the case of argon, there are significant differences in less highly-ionised stages. 

\begin{figure}
  \centering
  \includegraphics[width=0.7\textwidth]{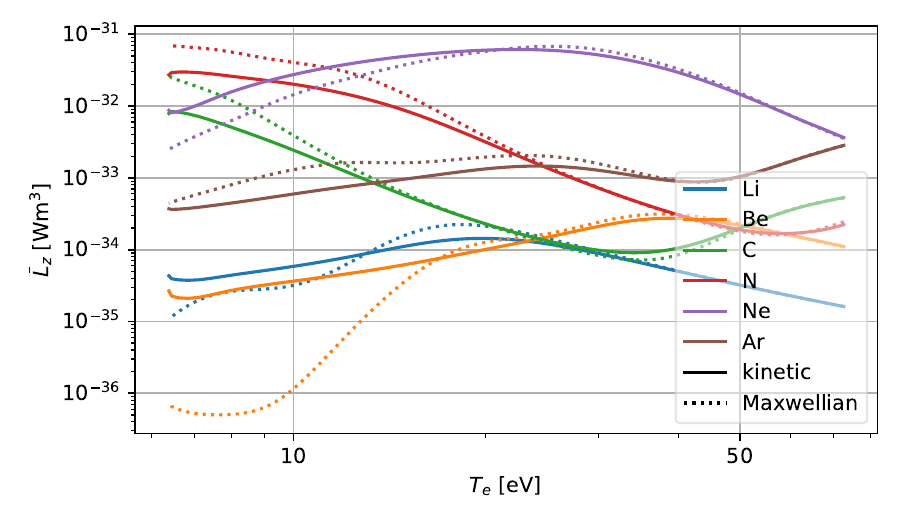}
  \caption{Average excitation radiation per ion (accounting for the ionisation balance) for all species considered.}
  \label{fig:eff_cooling_curves}
\end{figure}

The average or effective excitation radiation per ion, $\bar{L}_z$ (\ref{eq:eff_cooling_curve}), which is affected by both the radiative properties of each ionisation stage and the overall ionisation balance, is shown in Figure \ref{fig:eff_cooling_curves}. Moderate differences are seen at low $T_e$ across all species, with differences of over an order of magnitude in the case of beryllium. For all impurity species considered, the effect of non-local electron transport is to reduce the gradient of $\bar{L}_z$ with respect to $T_e$. 


Carrying out a similar analysis as for the ionisation balance at a single location (Figure \ref{fig:single_loc_iz_balance_Ne}) but for the radiative power loss due to excitation, $Q_{z}$ (\ref{eq:Q_z}), yields similar result - the flux-limited fluid model shows greater deviation than the fluid model, and the difference is greatest for a kinetic treatment of the electron transport with Maxwellian rates.


\begin{figure}
  \centering
  \begin{subfigure}{0.8\textwidth}
      \centering 
      \includegraphics[width=\textwidth]{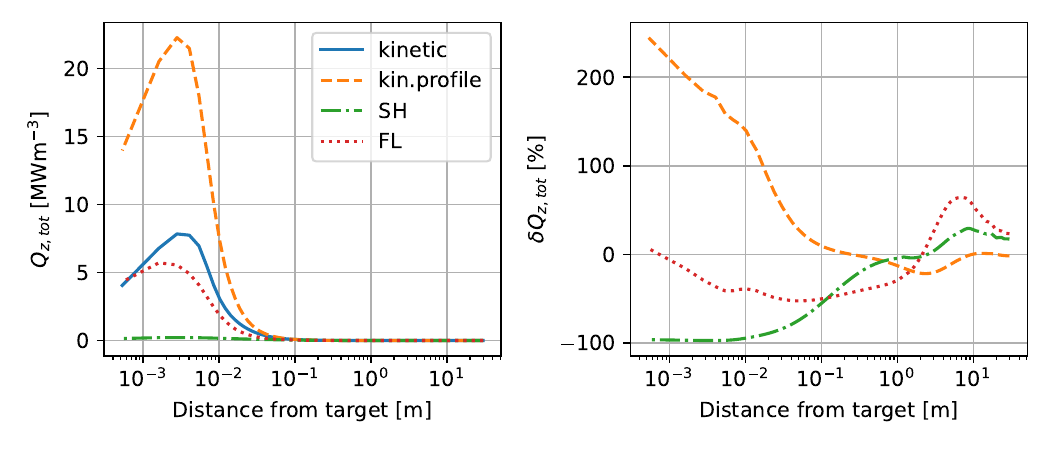}
      \caption{Carbon}
      \label{fig:Q_rad_error_C}
  \end{subfigure}
  \hfill
  \begin{subfigure}{0.8\textwidth}
      \centering 
      \includegraphics[width=\textwidth]{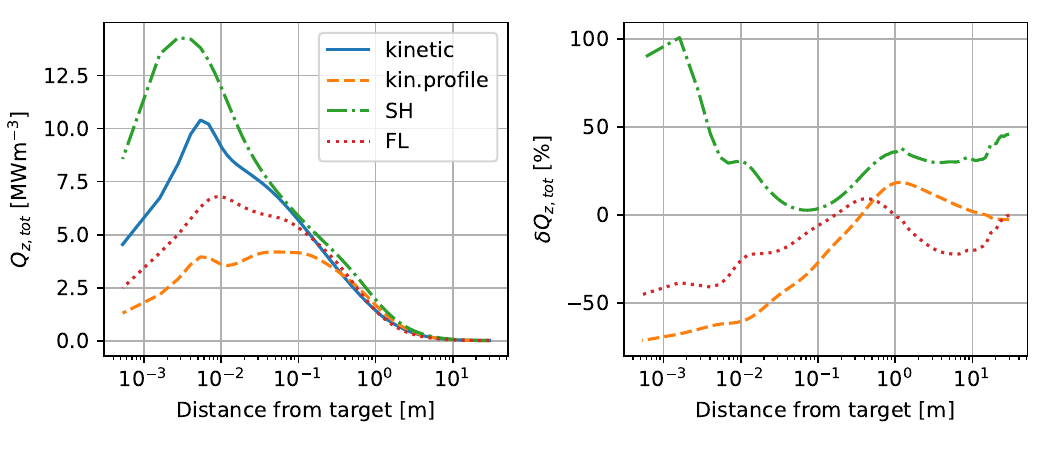}
      \caption{Neon}
      \label{fig:Q_rad_error_Ne}
  \end{subfigure}
  \hfill
  \begin{subfigure}{0.8\textwidth}
      \centering 
      \includegraphics[width=\textwidth]{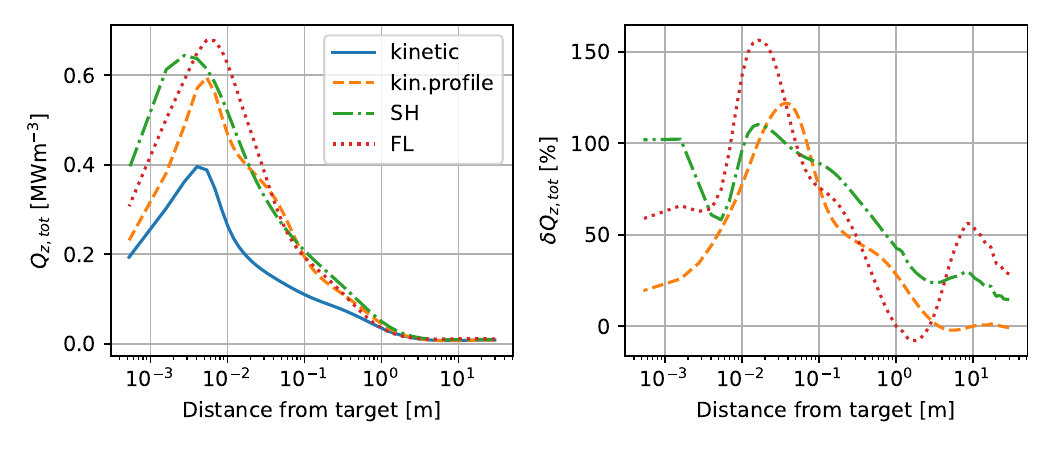}
      \caption{Argon}
      \label{fig:Q_rad_error_Ar}
  \end{subfigure}
  \caption{Total excitation radiation profiles (left) and errors compared to the kinetic calculation (right).}
  \label{fig:Q_rad_error}
\end{figure}

In Figure \ref{fig:Q_rad_error}, we plot the spatial profile of $Q_{z,tot}$\footnote{Note that $Q_{z,tot}$ provides information on the spatial profile of the radiation peak, but due to the fixed fraction impurity density assumption used here $Q_{z}$ is doubly affected by the $n_e$ profile. $\delta Q_{z,tot}$ is not, so it is heplful to show both quantities.} as well as the percentage deviation in $Q_{z,tot}$ compared to the fully kinetic treatment for carbon, neon and argon, i.e. $\delta Q_{z,tot} = (Q_{z,tot}^{\cdots} - Q_{z,tot}^{kinetic}) / Q_{z,tot}^{kinetic}$, where the ellipsis is one of `kin. profile', `SH' or `FL'.
Significant deviations are seen in both species, with both over- and underestimates of the radiative losses observed, particularly in the region close to the target.

\begin{figure}
  \centering
  \begin{subfigure}{0.8\textwidth}
      \centering 
      \includegraphics[width=\textwidth]{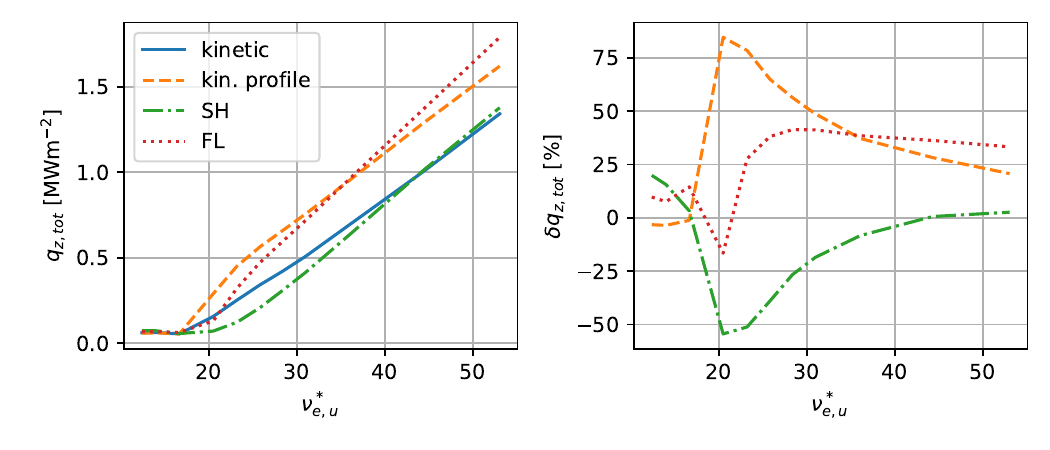}
      \caption{Carbon}
      \label{fig:q_z_tot_vs_nu_star_C}
  \end{subfigure}
  \hfill
  \begin{subfigure}{0.8\textwidth}
      \centering 
      \includegraphics[width=\textwidth]{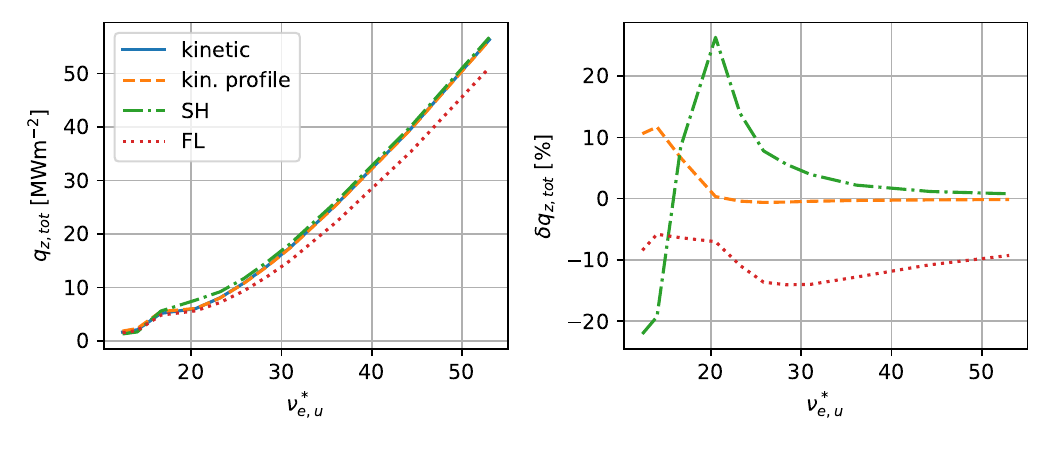}
      \caption{Neon}
      \label{fig:q_z_tot_vs_nu_star_Ne}
  \end{subfigure}
  \hfill
  \begin{subfigure}{0.8\textwidth}
      \centering 
      \includegraphics[width=\textwidth]{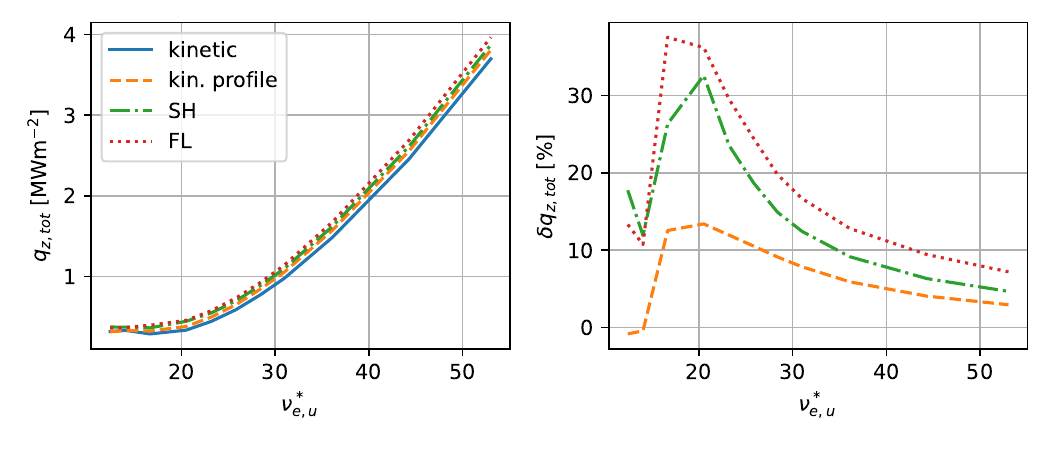}
      \caption{Argon}
      \label{fig:q_z_tot_vs_nu_star_Ar}
  \end{subfigure}
  \caption{Line-integrated radiated power from excitation (left) and error compared to a kinetic calculation (right). }
  \label{fig:q_z_tot_vs_nu_star}
\end{figure}

We can investigate the effect on the overall SOL power balance by comparing the line-integrated excitation radiation, $q_{z,tot}=\int_L Q_{z,tot}dx$. In the density scan described in Section \ref{sec:simulations}, Figure \ref{fig:q_z_tot_vs_nu_star} shows the value of $q_{z,tot}$ as a function of $\nu_{e,u}^*$ as well as the percentage deviation compared to the kinetic calculation, $\delta q_{z,tot} = (q_{z,tot}^{\cdots} - q_{z,tot}^{kinetic})/q_{z,tot}^{kinetic}$. Note that $q_{z,tot}$ can be higher than the input power to SOL in these simulations here because the plasma background is fixed. There are significant differences at intermediate collisionalities, albeit smaller than in the case where the spatial profiles were considered. 

\section{Discussion}
\label{sec:discussion}

Calculations of the effective ionisation rate coefficients show that assuming Maxwellian electrons can drastically underestimate the rates compared to a realistic electron distribution close to the target in SOL simulations, Figure \ref{fig:iz_coeffs}. This is particularly the case for highly-ionised impurity ions at low electron temperatures, while the rates for neutral and near-neutral particles show good agreement at all temperatures. The reason for the rate increases is the accumulation of fast electrons close to the target due to non-local transport.
The recombination rates are largely unaffected since thermalised electrons dominate the process, and so the resulting ionisation balance is altered.  
Since the enhancements to ionisation rates occur at low $T_e$, where the average impurity ionisation is low, the resultant modifications to the ionisation balance are more moderate than would appear at first glance from Figure \ref{fig:iz_coeffs}. Considering only the effect of the electron distribution, a kinetic treatment predicts higher mean charge by an amount, $\Delta \bar{z}$, which varies for the impurity species, but peaks at 0.56 in front of the target for carbon (Figure \ref{fig:avg_iz_and_error_C}), 1.56 for neon (Figure \ref{fig:avg_iz_and_error_Ne}) and 1.89 for argon (Figure \ref{fig:avg_iz_and_error_Ar}). 

\begin{figure}
  \centering
  \begin{subfigure}{0.49\textwidth}
      \centering 
      \includegraphics[width=\textwidth]{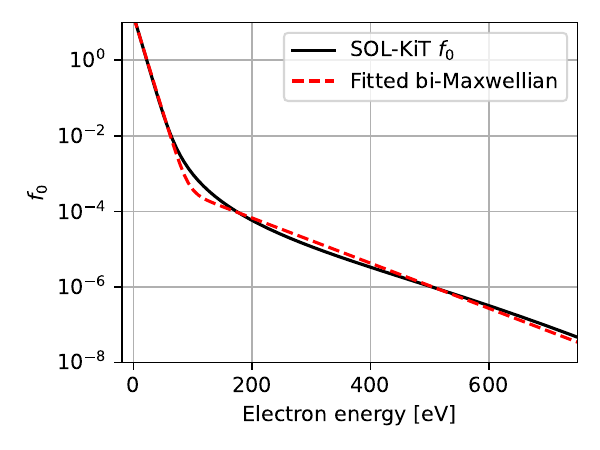}
      \caption{}
      \label{fig:fitted_bimaxwellian_example}
  \end{subfigure}
  \hfill
  \begin{subfigure}{0.49\textwidth}
      \centering 
      \includegraphics[width=\textwidth]{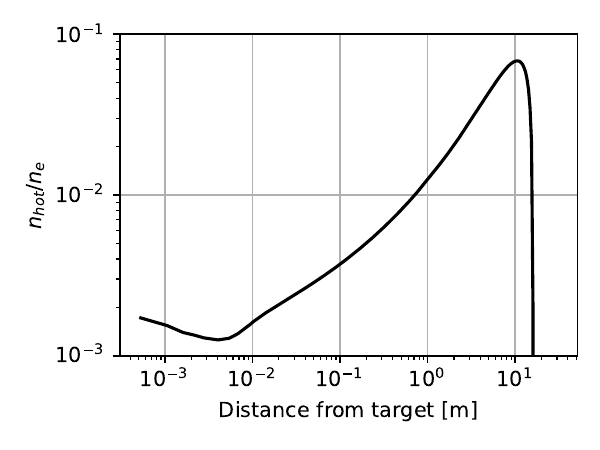}
      \caption{}
      \label{fig:bimaxwellian_fitted_hot_frac}
  \end{subfigure}
  \caption{(a) Isotropic part of electron distribution close to target as shown in Figure \protect{\ref{fig:impurity_runs_example_dist}} with a fitted bi-Maxwellian distribution. (b) Spatial profile of the fractional density of the hot population in fitted bi-Maxwellian distributions, which peaks at 7\%.}
  \label{fig:bimaxwellian_fitting}
\end{figure}

\begin{figure}
  \centering
  \begin{subfigure}{0.49\textwidth}
      \centering 
      \includegraphics[width=\textwidth]{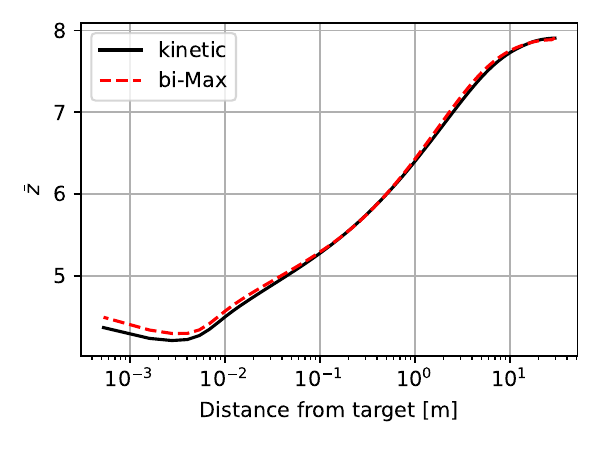}
      \caption{}
      \label{fig:bimax_Zavg_Ne}
  \end{subfigure}
  \hfill
  \begin{subfigure}{0.49\textwidth}
      \centering 
      \includegraphics[width=\textwidth]{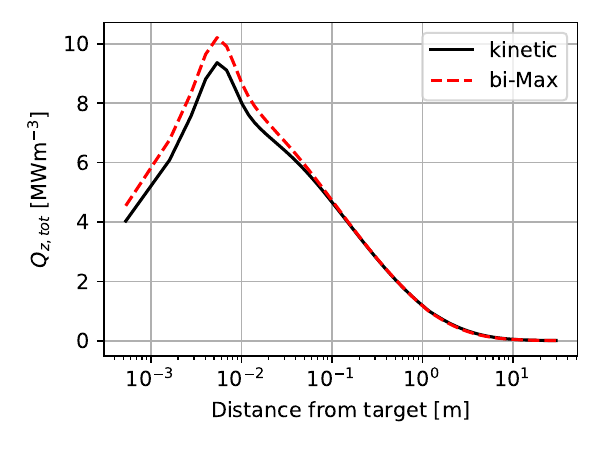}
      \caption{}
      \label{fig:bimax_Q_z_tot_Ne}
  \end{subfigure}
  \caption{Spatial profiles of (a) average ionisation and (b) total excitation radiation for neon calculated with fitted bi-Maxwellian electron distributions and compared to the kinetic treatment.}
  \label{fig:bimax_results}
\end{figure}

Similar results are seen in the energy loss rates due to excitation. 
There is an additional effect to consider however, because the total excitation radiation is affected by the excitation per ion, $L_z$, as well as the ionisation balance. 
The effect of the electron distribution alone leads to differences in the predicted excitation power loss profiles of up to 100-200\% (Figure \ref{fig:Q_rad_error}), but smaller differences in the line-integrated quantities (Figure \ref{fig:q_z_tot_vs_nu_star}).

The choice of electron transport model modifies the plasma profiles, and if Maxwellian electrons are assumed for the reaction rates then errors can be introduced. This is seen in Figures \ref{fig:single_loc_iz_balance_Ne}, where a flux-limiter applied to a fluid electron transport model produces worse agreement in the ionisation balance and radiative losses than a model with Spitzer-H\"{a}rm conductivity. In some cases the opposite is true, as seen in Figures \ref{fig:avg_iz_errors} and \ref{fig:Q_rad_error}, but it is dependent on the impurity species and spatial location being considered. Electron transport therefore can impact the atomic physics, and attempts to improve upon a Spitzer-H\"{a}rm treatment via flux limiters may reduce, rather than increase, accuracy of predicted quantities such as $Q_{z,tot}$. 

The study presented here is similar in nature to Zhao et al.\cite{Zhao2019}, where very good agreement was found between ADAS rates and a kinetic treatment of inelastic collision rates between electrons and neutral hydrogen in self-consistent simulations with the KIPP code. This finding has been replicated with SOL-KiT\cite{Mijin2020a,Power2023}. The findings presented here suggest this may not be the case for impurity species with higher atomic number or molecules.

The main limitations of this study are that a) only equilibrium SOL conditions have been considered, b) the presence of the impurity species was not treated self-consistently, i.e. the plasma background was fixed, and c) there is no impurity transport. On the one hand, transient SOL conditions may be expected to yield stronger kinetic effects\cite{Batishchev1999,Mijin2020a,Tskhakaya2009}. On the other hand, a self-consistent simulation would capture the depletion of the high-energy electron population from inelastic collisions with impurities, and so the effects seen here may be weakened. A self-consistent treatment with kinetic electrons and impurity transport would likely require novel techniques to compute the atomic kinetics without resorting to solving the full set of rate equations (\ref{eq:SIKE_dens_mat}) at each timestep. One viable approach may be to use the dressed cross-section model outlined by Tskhakaya\cite{tskhakaya_implementation_2023}. SIKE could be used to develop such cross-sections straightforwardly, perhaps parameterised in terms of the bi-Maxwellian character of the electron distribution (see below). 

\subsection{Bi-Maxwellian parameterisation}
\label{sec:bimaxwellian}
The kinetic treatment of the SOL electrons shows that hot electrons from upstream can stream collisionlessly along the flux tube toward the walls. The electron distribution near to the targets, such as the one shown in Figure \ref{fig:impurity_runs_example_dist}, would then appear to be well-described by a bi-Maxwellian made up of a bulk electron population and a minority hot fraction, i.e. $f_0(v) = f^{Max}_{hot} + f^{Max}_{bulk}$, where the hot population is a Maxwellian at the upstream electron temperature, $T_{e,u}$, and has a density $n_{hot}$, and the bulk population is a Maxwellian at the local electron temperature and has density $n_e-n_{hot}$. In Figure \ref{fig:bimaxwellian_fitting}, a fit of this type (where $n_{hot}$ is the only free parameter) to the $f_0$ from Figure \ref{fig:impurity_runs_example_dist} is shown, as well as a spatial profile of the fitted values of $n_{hot}$. It is then natural to ask whether such a distribution would accurately capture the impurity reaction rates seen in the results presented here. If this were true, it would then be a case of finding $n_{hot}$ to capture the rates more accurately, which may not be trivial but is much less computationally expensive than a fully kinetic treatment. This has been done for neon and is shown in Figure \ref{fig:bimax_results}, where there is good agreement between the results using fitted bi-Maxwellian distributions and the SOL-KiT distributions (compare with Figures \ref{fig:avg_iz_and_error_Ne} and \ref{fig:Q_rad_error_Ne}). This suggests a bi-Maxwellian treatment of impurity reaction rates may be a viable approach to improving their accuracy in regimes where non-local transport is important. 

\begin{figure}
    \centering 
    \includegraphics[width=0.6\textwidth]{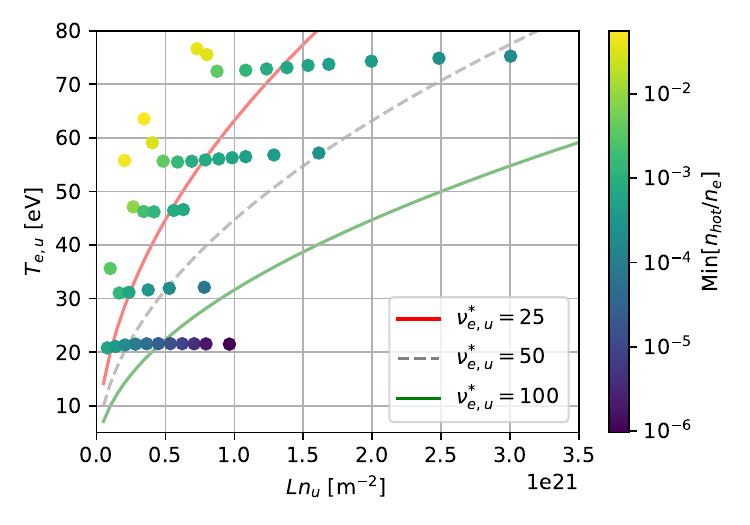}
    \caption{Minimum value of $n_{hot}/n_e$ in simulations at a range of collisionalities.}
    \label{fig:bimax_min_nhot_vs_nustar}
\end{figure}

This parameterisation of the electron distribution also provides a straightforward way of assessing the strength of any potential kinetic effects in different SOL conditions. In Figure \ref{fig:bimax_min_nhot_vs_nustar}, the minimum value of $n_{hot}/n_e$ in electron distributions in the divertor of the kinetic simulations in the study by Power et al.\cite{Power2023} is shown, plotted on the $Ln_u$-$T_{e,u}$ plane. The lowest values of $n_{hot}/n_e$ are around $10^{-6}$, seen in deeply detached conditions at very high collisionalities. While small, these values are non-negligible given the significant enhancement to rates seen in Figure \ref{fig:iz_coeffs}. As has been suggested previously\cite{Jayakumar1985}, the effects discussed here may be experimentally observable in spectroscopic studies of SOL plasmas, where absolute intensities and line emission ratios may be modified from predictions which assume Maxwellian electrons. Future work is planned to use this approach to identify specific enhanced spectral lines from impurities (or line emission ratios) in which to look for such effects.

\subsection{Consequences for detached plasmas}
As noted in Section \ref{sec:intro}, it has been suggested that kinetic effects are the cause of overestimated Langmuir probe temperature inferences in detached plasmas\cite{Batishchev1996,Batishchev1997}. It is therefore natural to question how the electron distributions of the kind seen here (Figure \ref{fig:impurity_runs_example_dist}) will impact the detached SOL. 

A (deeply) detached plasma can feature a cushion of neutral atoms and molecules. The interaction of tail electrons with the neutral atoms has been discussed here and elsewhere\cite{Batishchev1999,Mijin2019,Zhao2019}, but the impact on the hydrogenic molecular chemistry of a detached plasma is yet to be studied in detail. Hot electrons interacting with a dense molecular cloud can vibrationally excite the molecules, either directly or through electronic excitation and subsequent decay. Vibrationally excited molecules interact with the plasma, leading to additional dissociation and ion sinks through molecular activated dissociation / ionisation\cite{Verhaegh2023a}. These processes can be boosted significantly through a population of fast electrons in a $<1$ eV bulk electron regime, where the bulk electrons have insufficient energy to drive strong vibrational excitation (see, for example, the sensitivity of vibrational excitation with electron temperature in detailed molecular collisional-radiative modelling\cite{Kobussen2023}). Further collisional-radiative modelling studies, for example using the bi-maxwellian representation in Section \ref{sec:bimaxwellian}, are required to investigate the sensitivity of plasma-molecule interaction rates to kinetic effects.

The electronic excitation of molecules through an enhanced population of fast electrons would result in molecular Fulcher emission (595-640 nm) from excited molecules that can be diagnosed spectroscopically. Since the energy threshold (for ground-state molecules) for excitation to the Fulcher level is 13.9 eV\cite{Scarlett2021}, one would expect molecular Fulcher emission to be negligible for a (deeply) detached plasma without kinetic effects\cite{Verhaegh2023}. The presence of Fulcher emission in such a plasma could thus be used as a diagnostic for kinetic effects, which becomes more sensitive at lower bulk temperatures. For a hot electron population at $T_{hot} = 40$ eV, $n_{hot}/n_e$ values higher than $10^{-5}$, $10^{-4}$, $10^{-3}$ are expected to be observable at a bulk temperature $T_{e} < 0.2$ eV, $< 0.4$ eV, $< 1.1$ eV respectively. 

\section{Conclusions}
\label{sec:conclusion}

We have demonstrated that, in simulations of SOL plasmas exhibiting non-local parallel electron transport, there is a clear impact on electron-impurity reaction rates. The driver of this effect is the departure of electrons from a Maxwellian distribution, particularly close the divertor targets, where an enhanced population of fast electrons from upstream is present. The spatial extent of this effect is up to around 1m from the target in these simulations, but is strongest just a few cm away. It should be noted that such non-Maxwellian distributions are not a new phenomenon presented here, and have been seen elsewhere\cite{Batishchev1999,Tskhakaya2011,Tskhakaya2017,Chankin2018,Mijin2020a} The main limitation of this study is the fact that the plasma background was fixed and no impurity transport was considered, so it would be worthwhile investigating this effect further with a more self-consistent treatment of the plasma species present. In particular, the cooling effect of impurity radiation on the bulk plasma is critical in driving detachment, so it is important to identify whether the findings here effect that behaviour. 

Additionally, attempts to improve upon the Spitzer-H\"{a}rm electron thermal transport using a flux limiter (but still using Maxwellian-averaged rates for the electron-impurity reactions) may not improve, or may even decrease, simulation accuracy. 

We have shown that a bi-Maxwellian representation of the electron distributions, made up of a local thermal bulk and a hot population at the upstream temperature, appears to be a viable method of capturing the differences in reaction rates seen here. This offers a simple way of estimating the modification of electron-impact reaction rates close to the walls in SOL plasmas, and may offer a more computationally-tractable method of treating impurity atomic population kinetics.

Finally, we have discussed the possible consequences of kinetic effects in detached SOL regimes, and proposed a method of experimentally observing such effects.

\section*{Acknowledgements and supporting data}
This work was part funded by the UK Engineering and Physical Science Research Council (EPSRC) Energy Programme (grant numbers EP/W006839/1 and EP/R513052/1) and the UK Atomic Energy Authority (UKAEA). It has also been improved through informal discussions with colleagues at UKAEA and elsewhere, in particular David Moulton and Martin O'Mullane.

The simulation data (SIKE and SOL-KiT), analysis scripts and code used to generate the plots in this study can be found at \url{https://doi.org/10.14469/hpc/13959}. The simulations were carried out using the Imperial College London Research Computing Service (\url{https://doi.org/10.14469/hpc/2232}).

\bibliography{zotero_lib.bib}

\begin{thebibliography}{55}%
\makeatletter
\providecommand \@ifxundefined [1]{%
 \@ifx{#1\undefined}
}%
\providecommand \@ifnum [1]{%
 \ifnum #1\expandafter \@firstoftwo
 \else \expandafter \@secondoftwo
 \fi
}%
\providecommand \@ifx [1]{%
 \ifx #1\expandafter \@firstoftwo
 \else \expandafter \@secondoftwo
 \fi
}%
\providecommand \natexlab [1]{#1}%
\providecommand \enquote  [1]{``#1''}%
\providecommand \bibnamefont  [1]{#1}%
\providecommand \bibfnamefont [1]{#1}%
\providecommand \citenamefont [1]{#1}%
\providecommand \href@noop [0]{\@secondoftwo}%
\providecommand \href [0]{\begingroup \@sanitize@url \@href}%
\providecommand \@href[1]{\@@startlink{#1}\@@href}%
\providecommand \@@href[1]{\endgroup#1\@@endlink}%
\providecommand \@sanitize@url [0]{\catcode `\\12\catcode `\$12\catcode `\&12\catcode `\#12\catcode `\^12\catcode `\_12\catcode `\%12\relax}%
\providecommand \@@startlink[1]{}%
\providecommand \@@endlink[0]{}%
\providecommand \url  [0]{\begingroup\@sanitize@url \@url }%
\providecommand \@url [1]{\endgroup\@href {#1}{\urlprefix }}%
\providecommand \urlprefix  [0]{URL }%
\providecommand \Eprint [0]{\href }%
\providecommand \doibase [0]{http://dx.doi.org/}%
\providecommand \selectlanguage [0]{\@gobble}%
\providecommand \bibinfo  [0]{\@secondoftwo}%
\providecommand \bibfield  [0]{\@secondoftwo}%
\providecommand \translation [1]{[#1]}%
\providecommand \BibitemOpen [0]{}%
\providecommand \bibitemStop [0]{}%
\providecommand \bibitemNoStop [0]{.\EOS\space}%
\providecommand \EOS [0]{\spacefactor3000\relax}%
\providecommand \BibitemShut  [1]{\csname bibitem#1\endcsname}%
\let\auto@bib@innerbib\@empty
\bibitem [{\citenamefont {Sciortino}\ \emph {et~al.}(2021)\citenamefont {Sciortino}, \citenamefont {Odstrčil}, \citenamefont {Cavallaro}, \citenamefont {Smith}, \citenamefont {Meneghini}, \citenamefont {Reksoatmodjo}, \citenamefont {Linder}, \citenamefont {Lore}, \citenamefont {Howard}, \citenamefont {Marmar},\ and\ \citenamefont {Mordijck}}]{Sciortino2021}%
  \BibitemOpen
  \bibfield  {author} {\bibinfo {author} {\bibfnamefont {F.}~\bibnamefont {Sciortino}}, \bibinfo {author} {\bibfnamefont {T.}~\bibnamefont {Odstrčil}}, \bibinfo {author} {\bibfnamefont {A.}~\bibnamefont {Cavallaro}}, \bibinfo {author} {\bibfnamefont {S.}~\bibnamefont {Smith}}, \bibinfo {author} {\bibfnamefont {O.}~\bibnamefont {Meneghini}}, \bibinfo {author} {\bibfnamefont {R.}~\bibnamefont {Reksoatmodjo}}, \bibinfo {author} {\bibfnamefont {O.}~\bibnamefont {Linder}}, \bibinfo {author} {\bibfnamefont {J.~D.}\ \bibnamefont {Lore}}, \bibinfo {author} {\bibfnamefont {N.~T.}\ \bibnamefont {Howard}}, \bibinfo {author} {\bibfnamefont {E.~S.}\ \bibnamefont {Marmar}}, \ and\ \bibinfo {author} {\bibfnamefont {S.}~\bibnamefont {Mordijck}},\ }\bibfield  {title} {\enquote {\bibinfo {title} {Modeling of particle transport, neutrals and radiation in magnetically-confined plasmas with {AURORA}},}\ }\href {http://arxiv.org/abs/2106.04528} {\bibfield  {journal} {\bibinfo  {journal} {Plasma Physics and Controlled Fusion}\ ,\ \bibinfo {pages} {1--9}} (\bibinfo {year} {2021})}\BibitemShut {NoStop}%
\bibitem [{\citenamefont {Johnson}, \citenamefont {Loch},\ and\ \citenamefont {Ennis}(2019)}]{Johnson2019}%
  \BibitemOpen
  \bibfield  {author} {\bibinfo {author} {\bibfnamefont {C.~A.}\ \bibnamefont {Johnson}}, \bibinfo {author} {\bibfnamefont {S.~D.}\ \bibnamefont {Loch}}, \ and\ \bibinfo {author} {\bibfnamefont {D.~A.}\ \bibnamefont {Ennis}},\ }\bibfield  {title} {\enquote {\bibinfo {title} {{ColRadPy}: A python collisional radiative solver},}\ }\href {\doibase 10.1016/j.nme.2019.01.013} {\bibfield  {journal} {\bibinfo  {journal} {Nuclear Materials and Energy}\ }\textbf {\bibinfo {volume} {20}} (\bibinfo {year} {2019}),\ 10.1016/j.nme.2019.01.013}\BibitemShut {NoStop}%
\bibitem [{\citenamefont {Reiter}(2023)}]{Reiter2019}%
  \BibitemOpen
  \bibfield  {author} {\bibinfo {author} {\bibfnamefont {D.}~\bibnamefont {Reiter}},\ }\href {https://www.eirene.de/Documentation/eirene.pdf} {\enquote {\bibinfo {title} {The {EIRENE} code user manual manual version: April 28, 2023},}\ }\bibinfo {howpublished} {\url{https://www.eirene.de/Documentation/eirene.pdf}} (\bibinfo {year} {2023})\BibitemShut {NoStop}%
\bibitem [{\citenamefont {Kotov}\ \emph {et~al.}(2008)\citenamefont {Kotov}, \citenamefont {Reiter}, \citenamefont {Pitts}, \citenamefont {Jachmich}, \citenamefont {Huber},\ and\ \citenamefont {Coster}}]{Kotov2008}%
  \BibitemOpen
  \bibfield  {author} {\bibinfo {author} {\bibfnamefont {V.}~\bibnamefont {Kotov}}, \bibinfo {author} {\bibfnamefont {D.}~\bibnamefont {Reiter}}, \bibinfo {author} {\bibfnamefont {R.~A.}\ \bibnamefont {Pitts}}, \bibinfo {author} {\bibfnamefont {S.}~\bibnamefont {Jachmich}}, \bibinfo {author} {\bibfnamefont {A.}~\bibnamefont {Huber}}, \ and\ \bibinfo {author} {\bibfnamefont {D.~P.}\ \bibnamefont {Coster}},\ }\bibfield  {title} {\enquote {\bibinfo {title} {Numerical modelling of high density {JET} divertor plasma with the {SOLPS}4.2 (b2-{EIRENE}) code},}\ }\href {\doibase 10.1088/0741-3335/50/10/105012} {\bibfield  {journal} {\bibinfo  {journal} {Plasma Physics and Controlled Fusion}\ }\textbf {\bibinfo {volume} {50}} (\bibinfo {year} {2008}),\ 10.1088/0741-3335/50/10/105012}\BibitemShut {NoStop}%
\bibitem [{\citenamefont {Batishchev}\ \emph {et~al.}(1999)\citenamefont {Batishchev}, \citenamefont {Shoucri}, \citenamefont {Batishcheva},\ and\ \citenamefont {Shkarofsky}}]{Batishchev1999}%
  \BibitemOpen
  \bibfield  {author} {\bibinfo {author} {\bibfnamefont {O.~V.}\ \bibnamefont {Batishchev}}, \bibinfo {author} {\bibfnamefont {M.~M.}\ \bibnamefont {Shoucri}}, \bibinfo {author} {\bibfnamefont {A.~A.}\ \bibnamefont {Batishcheva}}, \ and\ \bibinfo {author} {\bibfnamefont {I.~P.}\ \bibnamefont {Shkarofsky}},\ }\bibfield  {title} {\enquote {\bibinfo {title} {Fully kinetic simulation of coupled plasma and neutral particles in scrape-off layer plasmas of fusion devices},}\ }\href {\doibase 10.1017/S0022377898007375} {\bibfield  {journal} {\bibinfo  {journal} {Journal of Plasma Physics}\ }\textbf {\bibinfo {volume} {61}},\ \bibinfo {pages} {347--364} (\bibinfo {year} {1999})}\BibitemShut {NoStop}%
\bibitem [{\citenamefont {Tskhakaya}\ \emph {et~al.}(2009)\citenamefont {Tskhakaya}, \citenamefont {Pitts}, \citenamefont {Fundamenski}, \citenamefont {Eich}, \citenamefont {Kuhn},\ and\ \citenamefont {{JET EFDA Contributors}}}]{Tskhakaya2009}%
  \BibitemOpen
  \bibfield  {author} {\bibinfo {author} {\bibfnamefont {D.}~\bibnamefont {Tskhakaya}}, \bibinfo {author} {\bibfnamefont {R.~A.}\ \bibnamefont {Pitts}}, \bibinfo {author} {\bibfnamefont {W.}~\bibnamefont {Fundamenski}}, \bibinfo {author} {\bibfnamefont {T.}~\bibnamefont {Eich}}, \bibinfo {author} {\bibfnamefont {S.}~\bibnamefont {Kuhn}}, \ and\ \bibinfo {author} {\bibnamefont {{JET EFDA Contributors}}},\ }\bibfield  {title} {\enquote {\bibinfo {title} {Kinetic simulations of the parallel transport in the {JET} scrape-off layer},}\ }\href {\doibase 10.1016/j.jnucmat.2009.01.150} {\bibfield  {journal} {\bibinfo  {journal} {Journal of Nuclear Materials}\ }\textbf {\bibinfo {volume} {390-391}},\ \bibinfo {pages} {335--338} (\bibinfo {year} {2009})},\ \bibinfo {note} {publisher: Elsevier B.V.}\BibitemShut {Stop}%
\bibitem [{\citenamefont {Chankin}, \citenamefont {Corrigan},\ and\ \citenamefont {Jaervinen}(2018)}]{Chankin2018}%
  \BibitemOpen
  \bibfield  {author} {\bibinfo {author} {\bibfnamefont {A.~V.}\ \bibnamefont {Chankin}}, \bibinfo {author} {\bibfnamefont {G.}~\bibnamefont {Corrigan}}, \ and\ \bibinfo {author} {\bibfnamefont {A.~E.}\ \bibnamefont {Jaervinen}},\ }\bibfield  {title} {\enquote {\bibinfo {title} {Assessment of the strength of kinetic effects of parallel electron transport in the {SOL} and divertor of {JET} high radiative h-mode plasmas using {EDGE}2d-{EIRENE} and {KIPP} codes},}\ }\href {\doibase 10.1088/1361-6587/aae0a0} {\bibfield  {journal} {\bibinfo  {journal} {Plasma Physics and Controlled Fusion}\ }\textbf {\bibinfo {volume} {60}} (\bibinfo {year} {2018}),\ 10.1088/1361-6587/aae0a0},\ \bibinfo {note} {publisher: {IOP} Publishing}\BibitemShut {NoStop}%
\bibitem [{\citenamefont {Mijin}\ \emph {et~al.}(2020)\citenamefont {Mijin}, \citenamefont {Militello}, \citenamefont {Newton}, \citenamefont {Omotani},\ and\ \citenamefont {Kingham}}]{Mijin2020a}%
  \BibitemOpen
  \bibfield  {author} {\bibinfo {author} {\bibfnamefont {S.}~\bibnamefont {Mijin}}, \bibinfo {author} {\bibfnamefont {F.}~\bibnamefont {Militello}}, \bibinfo {author} {\bibfnamefont {S.}~\bibnamefont {Newton}}, \bibinfo {author} {\bibfnamefont {J.}~\bibnamefont {Omotani}}, \ and\ \bibinfo {author} {\bibfnamefont {R.~J.}\ \bibnamefont {Kingham}},\ }\bibfield  {title} {\enquote {\bibinfo {title} {Kinetic and fluid simulations of parallel electron transport during equilibria and transients in the scrape-off layer},}\ }\href {\doibase 10.1088/1361-6587/ab9b39} {\bibfield  {journal} {\bibinfo  {journal} {Plasma Physics and Controlled Fusion}\ }\textbf {\bibinfo {volume} {62}},\ \bibinfo {pages} {095004} (\bibinfo {year} {2020})}\BibitemShut {NoStop}%
\bibitem [{\citenamefont {Wigram}\ \emph {et~al.}(2020)\citenamefont {Wigram}, \citenamefont {Ridgers}, \citenamefont {Dudson}, \citenamefont {Brodrick},\ and\ \citenamefont {Omotani}}]{Wigram2020}%
  \BibitemOpen
  \bibfield  {author} {\bibinfo {author} {\bibfnamefont {M.}~\bibnamefont {Wigram}}, \bibinfo {author} {\bibfnamefont {C.~P.}\ \bibnamefont {Ridgers}}, \bibinfo {author} {\bibfnamefont {B.}~\bibnamefont {Dudson}}, \bibinfo {author} {\bibfnamefont {J.~P.}\ \bibnamefont {Brodrick}}, \ and\ \bibinfo {author} {\bibfnamefont {J.~T.}\ \bibnamefont {Omotani}},\ }\bibfield  {title} {\enquote {\bibinfo {title} {Incorporating nonlocal parallel thermal transport in 1d {ITER} {SOL} modelling},}\ }\href@noop {} {\  (\bibinfo {year} {2020})}\BibitemShut {NoStop}%
\bibitem [{\citenamefont {Power}\ \emph {et~al.}(2023)\citenamefont {Power}, \citenamefont {Mijin}, \citenamefont {Wigram}, \citenamefont {Militello},\ and\ \citenamefont {Kingham}}]{Power2023}%
  \BibitemOpen
  \bibfield  {author} {\bibinfo {author} {\bibfnamefont {D.}~\bibnamefont {Power}}, \bibinfo {author} {\bibfnamefont {S.}~\bibnamefont {Mijin}}, \bibinfo {author} {\bibfnamefont {M.}~\bibnamefont {Wigram}}, \bibinfo {author} {\bibfnamefont {F.}~\bibnamefont {Militello}}, \ and\ \bibinfo {author} {\bibfnamefont {R.}~\bibnamefont {Kingham}},\ }\bibfield  {title} {\enquote {\bibinfo {title} {Scaling laws for electron kinetic effects in tokamak scrape-off layer plasmas},}\ }\href {\doibase 10.1088/1741-4326/acdca6} {\bibfield  {journal} {\bibinfo  {journal} {Nuclear Fusion}\ }\textbf {\bibinfo {volume} {63}},\ \bibinfo {pages} {086013} (\bibinfo {year} {2023})},\ \Eprint {http://arxiv.org/abs/2208.10862} {2208.10862} \BibitemShut {NoStop}%
\bibitem [{\citenamefont {Veselova}\ \emph {et~al.}(2021)\citenamefont {Veselova}, \citenamefont {Kaveeva}, \citenamefont {Rozhansky}, \citenamefont {Senichenkov}, \citenamefont {Poletaeva}, \citenamefont {Pitts},\ and\ \citenamefont {Bonnin}}]{Veselova2021}%
  \BibitemOpen
  \bibfield  {author} {\bibinfo {author} {\bibfnamefont {I.}~\bibnamefont {Veselova}}, \bibinfo {author} {\bibfnamefont {E.}~\bibnamefont {Kaveeva}}, \bibinfo {author} {\bibfnamefont {V.}~\bibnamefont {Rozhansky}}, \bibinfo {author} {\bibfnamefont {I.}~\bibnamefont {Senichenkov}}, \bibinfo {author} {\bibfnamefont {A.}~\bibnamefont {Poletaeva}}, \bibinfo {author} {\bibfnamefont {R.~A.}\ \bibnamefont {Pitts}}, \ and\ \bibinfo {author} {\bibfnamefont {X.}~\bibnamefont {Bonnin}},\ }\bibfield  {title} {\enquote {\bibinfo {title} {{SOLPS}-{ITER} drift modelling of {ITER} burning plasmas with narrow near-{SOL} heat flux channels},}\ }\href {\doibase 10.1016/j.nme.2020.100870} {\bibfield  {journal} {\bibinfo  {journal} {Nuclear Materials and Energy}\ }\textbf {\bibinfo {volume} {26}},\ \bibinfo {pages} {100870} (\bibinfo {year} {2021})},\ \bibinfo {note} {publisher: Elsevier Ltd}\BibitemShut {NoStop}%
\bibitem [{\citenamefont {Rubino}\ \emph {et~al.}(2017)\citenamefont {Rubino}, \citenamefont {Ambrosino}, \citenamefont {Calabrò}, \citenamefont {Pericoli~Ridolfini},\ and\ \citenamefont {Viola}}]{Rubino2017}%
  \BibitemOpen
  \bibfield  {author} {\bibinfo {author} {\bibfnamefont {G.}~\bibnamefont {Rubino}}, \bibinfo {author} {\bibfnamefont {R.}~\bibnamefont {Ambrosino}}, \bibinfo {author} {\bibfnamefont {G.}~\bibnamefont {Calabrò}}, \bibinfo {author} {\bibfnamefont {V.}~\bibnamefont {Pericoli~Ridolfini}}, \ and\ \bibinfo {author} {\bibfnamefont {B.}~\bibnamefont {Viola}},\ }\bibfield  {title} {\enquote {\bibinfo {title} {Comparative analysis of the {SOL} plasma in {DEMO} using {EDGE}2d/{EIRENE} and {TECXY} codes},}\ }\href {\doibase 10.1016/j.nme.2016.11.004} {\bibfield  {journal} {\bibinfo  {journal} {Nuclear Materials and Energy}\ }\textbf {\bibinfo {volume} {12}},\ \bibinfo {pages} {864--868} (\bibinfo {year} {2017})},\ \bibinfo {note} {publisher: Elsevier Ltd}\BibitemShut {NoStop}%
\bibitem [{\citenamefont {Chankin}\ and\ \citenamefont {Coster}(2009)}]{Chankin2009a}%
  \BibitemOpen
  \bibfield  {author} {\bibinfo {author} {\bibfnamefont {A.~V.}\ \bibnamefont {Chankin}}\ and\ \bibinfo {author} {\bibfnamefont {D.~P.}\ \bibnamefont {Coster}},\ }\bibfield  {title} {\enquote {\bibinfo {title} {Comparison of 2d models for the plasma edge with experimental measurements and assessment of deficiencies},}\ }\href {\doibase 10.1016/j.jnucmat.2009.01.307} {\bibfield  {journal} {\bibinfo  {journal} {Journal of Nuclear Materials}\ }\textbf {\bibinfo {volume} {390-391}},\ \bibinfo {pages} {319--324} (\bibinfo {year} {2009})},\ \bibinfo {note} {publisher: Elsevier B.V.}\BibitemShut {Stop}%
\bibitem [{\citenamefont {Batishchev}\ \emph {et~al.}(1997)\citenamefont {Batishchev}, \citenamefont {Krasheninnikov}, \citenamefont {Catto}, \citenamefont {Batishcheva}, \citenamefont {Sigmar}, \citenamefont {Xu}, \citenamefont {Byers}, \citenamefont {Rognlien}, \citenamefont {Cohen}, \citenamefont {Shoucri},\ and\ \citenamefont {Shkarofskii}}]{Batishchev1997}%
  \BibitemOpen
  \bibfield  {author} {\bibinfo {author} {\bibfnamefont {O.~V.}\ \bibnamefont {Batishchev}}, \bibinfo {author} {\bibfnamefont {S.~I.}\ \bibnamefont {Krasheninnikov}}, \bibinfo {author} {\bibfnamefont {P.~J.}\ \bibnamefont {Catto}}, \bibinfo {author} {\bibfnamefont {A.~A.}\ \bibnamefont {Batishcheva}}, \bibinfo {author} {\bibfnamefont {D.~J.}\ \bibnamefont {Sigmar}}, \bibinfo {author} {\bibfnamefont {X.~Q.}\ \bibnamefont {Xu}}, \bibinfo {author} {\bibfnamefont {J.~A.}\ \bibnamefont {Byers}}, \bibinfo {author} {\bibfnamefont {T.~D.}\ \bibnamefont {Rognlien}}, \bibinfo {author} {\bibfnamefont {R.~H.}\ \bibnamefont {Cohen}}, \bibinfo {author} {\bibfnamefont {M.~M.}\ \bibnamefont {Shoucri}}, \ and\ \bibinfo {author} {\bibfnamefont {I.~P.}\ \bibnamefont {Shkarofskii}},\ }\bibfield  {title} {\enquote {\bibinfo {title} {Kinetic effects in tokamak scrape-off layer plasmas},}\ }\href {\doibase 10.1063/1.872280} {\bibfield  {journal} {\bibinfo  {journal} {Physics of Plasmas}\ }\textbf {\bibinfo {volume} {4}},\ \bibinfo {pages} {1672--1680} (\bibinfo {year} {1997})}\BibitemShut {NoStop}%
\bibitem [{\citenamefont {Verhaegh}\ \emph {et~al.}(2017)\citenamefont {Verhaegh}, \citenamefont {Lipschultz}, \citenamefont {Duval}, \citenamefont {Harrison}, \citenamefont {Reimerdes}, \citenamefont {Theiler}, \citenamefont {Labit}, \citenamefont {Maurizio}, \citenamefont {Marini}, \citenamefont {Nespoli}, \citenamefont {Sheikh}, \citenamefont {Tsui}, \citenamefont {Vianello},\ and\ \citenamefont {Vijvers}}]{Verhaegh2017}%
  \BibitemOpen
  \bibfield  {author} {\bibinfo {author} {\bibfnamefont {K.}~\bibnamefont {Verhaegh}}, \bibinfo {author} {\bibfnamefont {B.}~\bibnamefont {Lipschultz}}, \bibinfo {author} {\bibfnamefont {B.~P.}\ \bibnamefont {Duval}}, \bibinfo {author} {\bibfnamefont {J.~R.}\ \bibnamefont {Harrison}}, \bibinfo {author} {\bibfnamefont {H.}~\bibnamefont {Reimerdes}}, \bibinfo {author} {\bibfnamefont {C.}~\bibnamefont {Theiler}}, \bibinfo {author} {\bibfnamefont {B.}~\bibnamefont {Labit}}, \bibinfo {author} {\bibfnamefont {R.}~\bibnamefont {Maurizio}}, \bibinfo {author} {\bibfnamefont {C.}~\bibnamefont {Marini}}, \bibinfo {author} {\bibfnamefont {F.}~\bibnamefont {Nespoli}}, \bibinfo {author} {\bibfnamefont {U.}~\bibnamefont {Sheikh}}, \bibinfo {author} {\bibfnamefont {C.~K.}\ \bibnamefont {Tsui}}, \bibinfo {author} {\bibfnamefont {N.}~\bibnamefont {Vianello}}, \ and\ \bibinfo {author} {\bibfnamefont {W.~A.}\ \bibnamefont {Vijvers}},\ }\bibfield  {title} {\enquote {\bibinfo {title} {Spectroscopic investigations of divertor detachment in {TCV}},}\ }\href {\doibase 10.1016/j.nme.2017.01.004} {\bibfield  {journal} {\bibinfo  {journal} {Nuclear Materials and Energy}\ }\textbf {\bibinfo {volume} {12}},\ \bibinfo {pages} {1112--1117} (\bibinfo {year} {2017})},\ \bibinfo {note} {publisher: Elsevier Ltd},\ \Eprint {http://arxiv.org/abs/1607.04539} {1607.04539} \BibitemShut {NoStop}%
\bibitem [{\citenamefont {Batishchev}\ \emph {et~al.}(1996)\citenamefont {Batishchev}, \citenamefont {Xu}, \citenamefont {Byers}, \citenamefont {Cohen}, \citenamefont {Krasheninnikov}, \citenamefont {Rognlien},\ and\ \citenamefont {Sigmar}}]{Batishchev1996}%
  \BibitemOpen
  \bibfield  {author} {\bibinfo {author} {\bibfnamefont {O.~V.}\ \bibnamefont {Batishchev}}, \bibinfo {author} {\bibfnamefont {X.~Q.}\ \bibnamefont {Xu}}, \bibinfo {author} {\bibfnamefont {J.~A.}\ \bibnamefont {Byers}}, \bibinfo {author} {\bibfnamefont {R.~H.}\ \bibnamefont {Cohen}}, \bibinfo {author} {\bibfnamefont {S.~I.}\ \bibnamefont {Krasheninnikov}}, \bibinfo {author} {\bibfnamefont {T.~D.}\ \bibnamefont {Rognlien}}, \ and\ \bibinfo {author} {\bibfnamefont {D.~J.}\ \bibnamefont {Sigmar}},\ }\bibfield  {title} {\enquote {\bibinfo {title} {Kinetic effects on particle and heat fluxes in detached plasmas},}\ }\href {\doibase 10.1063/1.871615} {\bibfield  {journal} {\bibinfo  {journal} {Physics of Plasmas}\ }\textbf {\bibinfo {volume} {3}},\ \bibinfo {pages} {3386--3396} (\bibinfo {year} {1996})}\BibitemShut {NoStop}%
\bibitem [{\citenamefont {Fundamenski}(2005)}]{Fundamenski2005}%
  \BibitemOpen
  \bibfield  {author} {\bibinfo {author} {\bibfnamefont {W.}~\bibnamefont {Fundamenski}},\ }\bibfield  {title} {\enquote {\bibinfo {title} {Parallel heat flux limits in the tokamak scrape-off layer},}\ }\href {\doibase 10.1088/0741-3335/47/11/R01} {\bibfield  {journal} {\bibinfo  {journal} {Plasma Physics and Controlled Fusion}\ }\textbf {\bibinfo {volume} {47}} (\bibinfo {year} {2005}),\ 10.1088/0741-3335/47/11/R01}\BibitemShut {NoStop}%
\bibitem [{\citenamefont {Brodrick}\ \emph {et~al.}(9 01)\citenamefont {Brodrick}, \citenamefont {Kingham}, \citenamefont {Marinak}, \citenamefont {Patel}, \citenamefont {Chankin}, \citenamefont {Omotani}, \citenamefont {Umansky}, \citenamefont {Del~Sorbo}, \citenamefont {Dudson}, \citenamefont {Parker}, \citenamefont {Kerbel}, \citenamefont {Sherlock},\ and\ \citenamefont {Ridgers}}]{Brodrick2017}%
  \BibitemOpen
  \bibfield  {author} {\bibinfo {author} {\bibfnamefont {J.~P.}\ \bibnamefont {Brodrick}}, \bibinfo {author} {\bibfnamefont {R.~J.}\ \bibnamefont {Kingham}}, \bibinfo {author} {\bibfnamefont {M.~M.}\ \bibnamefont {Marinak}}, \bibinfo {author} {\bibfnamefont {M.~V.}\ \bibnamefont {Patel}}, \bibinfo {author} {\bibfnamefont {A.~V.}\ \bibnamefont {Chankin}}, \bibinfo {author} {\bibfnamefont {J.~T.}\ \bibnamefont {Omotani}}, \bibinfo {author} {\bibfnamefont {M.~V.}\ \bibnamefont {Umansky}}, \bibinfo {author} {\bibfnamefont {D.}~\bibnamefont {Del~Sorbo}}, \bibinfo {author} {\bibfnamefont {B.}~\bibnamefont {Dudson}}, \bibinfo {author} {\bibfnamefont {J.~T.}\ \bibnamefont {Parker}}, \bibinfo {author} {\bibfnamefont {G.~D.}\ \bibnamefont {Kerbel}}, \bibinfo {author} {\bibfnamefont {M.}~\bibnamefont {Sherlock}}, \ and\ \bibinfo {author} {\bibfnamefont {C.~P.}\ \bibnamefont {Ridgers}},\ }\bibfield  {title} {\enquote {\bibinfo {title} {Testing nonlocal models of electron thermal conduction for magnetic and inertial confinement fusion applications},}\ }\href {\doibase 10.1063/1.5001079} {\bibfield  {journal} {\bibinfo  {journal} {Physics of Plasmas}\ }\textbf {\bibinfo {volume} {24}} (\bibinfo {year} {2017-09-01}),\ 10.1063/1.5001079},\ \bibinfo {note} {publisher: American Institute of Physics Inc.},\ \Eprint {http://arxiv.org/abs/1704.08963} {1704.08963} \BibitemShut {NoStop}%
\bibitem [{\citenamefont {Coster}(2011)}]{Coster2011}%
  \BibitemOpen
  \bibfield  {author} {\bibinfo {author} {\bibfnamefont {D.~P.}\ \bibnamefont {Coster}},\ }\bibfield  {title} {\enquote {\bibinfo {title} {Detachment physics in {SOLPS} simulations},}\ }\href {\doibase 10.1016/j.jnucmat.2010.12.223} {\bibfield  {journal} {\bibinfo  {journal} {Journal of Nuclear Materials}\ }\textbf {\bibinfo {volume} {415}},\ \bibinfo {pages} {S545--S548} (\bibinfo {year} {2011})},\ \bibinfo {note} {publisher: Elsevier B.V.}\BibitemShut {Stop}%
\bibitem [{\citenamefont {Garland}\ \emph {et~al.}(2020)\citenamefont {Garland}, \citenamefont {Chung}, \citenamefont {Fontes}, \citenamefont {Zammit}, \citenamefont {Colgan}, \citenamefont {Elder}, \citenamefont {{McDevitt}}, \citenamefont {Wildey},\ and\ \citenamefont {Tang}}]{Garland2020}%
  \BibitemOpen
  \bibfield  {author} {\bibinfo {author} {\bibfnamefont {N.~A.}\ \bibnamefont {Garland}}, \bibinfo {author} {\bibfnamefont {H.~K.}\ \bibnamefont {Chung}}, \bibinfo {author} {\bibfnamefont {C.~J.}\ \bibnamefont {Fontes}}, \bibinfo {author} {\bibfnamefont {M.~C.}\ \bibnamefont {Zammit}}, \bibinfo {author} {\bibfnamefont {J.}~\bibnamefont {Colgan}}, \bibinfo {author} {\bibfnamefont {T.}~\bibnamefont {Elder}}, \bibinfo {author} {\bibfnamefont {C.~J.}\ \bibnamefont {{McDevitt}}}, \bibinfo {author} {\bibfnamefont {T.~M.}\ \bibnamefont {Wildey}}, \ and\ \bibinfo {author} {\bibfnamefont {X.~Z.}\ \bibnamefont {Tang}},\ }\bibfield  {title} {\enquote {\bibinfo {title} {Impact of a minority relativistic electron tail interacting with a thermal plasma containing high-atomic-number impurities},}\ }\href {\doibase 10.1063/5.0003638} {\bibfield  {journal} {\bibinfo  {journal} {Physics of Plasmas}\ }\textbf {\bibinfo {volume} {27}} (\bibinfo {year} {2020}),\ 10.1063/5.0003638},\ \bibinfo {note} {publisher: {AIP} Publishing {LLC}}\BibitemShut {NoStop}%
\bibitem [{\citenamefont {Garland}\ \emph {et~al.}(2022)\citenamefont {Garland}, \citenamefont {Chung}, \citenamefont {Zammit}, \citenamefont {{McDevitt}}, \citenamefont {Colgan}, \citenamefont {Fontes},\ and\ \citenamefont {Tang}}]{Garland2022}%
  \BibitemOpen
  \bibfield  {author} {\bibinfo {author} {\bibfnamefont {N.~A.}\ \bibnamefont {Garland}}, \bibinfo {author} {\bibfnamefont {H.~K.}\ \bibnamefont {Chung}}, \bibinfo {author} {\bibfnamefont {M.~C.}\ \bibnamefont {Zammit}}, \bibinfo {author} {\bibfnamefont {C.~J.}\ \bibnamefont {{McDevitt}}}, \bibinfo {author} {\bibfnamefont {J.}~\bibnamefont {Colgan}}, \bibinfo {author} {\bibfnamefont {C.~J.}\ \bibnamefont {Fontes}}, \ and\ \bibinfo {author} {\bibfnamefont {X.~Z.}\ \bibnamefont {Tang}},\ }\bibfield  {title} {\enquote {\bibinfo {title} {Understanding how minority relativistic electron populations may dominate charge state balance and radiative cooling of a post-thermal quench tokamak plasma},}\ }\href {\doibase 10.1063/5.0071996} {\bibfield  {journal} {\bibinfo  {journal} {Physics of Plasmas}\ }\textbf {\bibinfo {volume} {29}} (\bibinfo {year} {2022}),\ 10.1063/5.0071996},\ \bibinfo {note} {publisher: {AIP} Publishing {LLC}}\BibitemShut {NoStop}%
\bibitem [{\citenamefont {Smith}(2003)}]{smith_enhancement_2003}%
  \BibitemOpen
  \bibfield  {author} {\bibinfo {author} {\bibfnamefont {G.~R.}\ \bibnamefont {Smith}},\ }\bibfield  {title} {\enquote {\bibinfo {title} {Enhancement of the helium resonance lines in the solar atmosphere by suprathermal electron excitation - {II}. {Non}-{Maxwellian} electron distributions},}\ }\href {https://academic.oup.com/mnras/article/341/1/143/997142} {\bibfield  {journal} {\bibinfo  {journal} {Monthly Notices of the Royal Astronomical Society}\ }\textbf {\bibinfo {volume} {341}},\ \bibinfo {pages} {143--163} (\bibinfo {year} {2003})}\BibitemShut {NoStop}%
\bibitem [{\citenamefont {Dudík}\ \emph {et~al.}(2003)\citenamefont {Dudík}, \citenamefont {Dzifčáková}, \citenamefont {Meyer-Vernet}, \citenamefont {Del~Zanna}, \citenamefont {Young}, \citenamefont {Giunta}, \citenamefont {Sylwester}, \citenamefont {Sylwester}, \citenamefont {Oka}, \citenamefont {Mason}, \citenamefont {Vocks}, \citenamefont {Matteini}, \citenamefont {Krucker}, \citenamefont {Williams},\ and\ \citenamefont {Mackovjak}}]{dudik_nonequilibrium_2017}%
  \BibitemOpen
  \bibfield  {author} {\bibinfo {author} {\bibfnamefont {J.}~\bibnamefont {Dudík}}, \bibinfo {author} {\bibfnamefont {E.}~\bibnamefont {Dzifčáková}}, \bibinfo {author} {\bibfnamefont {N.}~\bibnamefont {Meyer-Vernet}}, \bibinfo {author} {\bibfnamefont {G.}~\bibnamefont {Del~Zanna}}, \bibinfo {author} {\bibfnamefont {P.~R.}\ \bibnamefont {Young}}, \bibinfo {author} {\bibfnamefont {A.}~\bibnamefont {Giunta}}, \bibinfo {author} {\bibfnamefont {B.}~\bibnamefont {Sylwester}}, \bibinfo {author} {\bibfnamefont {J.}~\bibnamefont {Sylwester}}, \bibinfo {author} {\bibfnamefont {M.}~\bibnamefont {Oka}}, \bibinfo {author} {\bibfnamefont {H.~E.}\ \bibnamefont {Mason}}, \bibinfo {author} {\bibfnamefont {C.}~\bibnamefont {Vocks}}, \bibinfo {author} {\bibfnamefont {L.}~\bibnamefont {Matteini}}, \bibinfo {author} {\bibfnamefont {S.}~\bibnamefont {Krucker}}, \bibinfo {author} {\bibfnamefont {D.~R.}\ \bibnamefont {Williams}}, \ and\ \bibinfo {author} {\bibfnamefont {S.}~\bibnamefont {Mackovjak}},\ }\bibfield  {title} {\enquote {\bibinfo {title} {Nonequilibrium {Processes} in the {Solar} {Corona}, {Transition} {Region}, {Flares}, and {Solar} {Wind} ({Invited} {Review})},}\ }\href {\doibase 10.1007/s11207-017-1125-0} {\bibfield  {journal} {\bibinfo  {journal} {Solar Physics}\ }\textbf {\bibinfo {volume} {292}},\ \bibinfo {pages} {100} (\bibinfo {year} {2003})}\BibitemShut {NoStop}%
\bibitem [{\citenamefont {Chung}\ \emph {et~al.}(2005)\citenamefont {Chung}, \citenamefont {Chen}, \citenamefont {Morgan}, \citenamefont {Ralchenko},\ and\ \citenamefont {Lee}}]{Chung2005}%
  \BibitemOpen
  \bibfield  {author} {\bibinfo {author} {\bibfnamefont {H.~K.}\ \bibnamefont {Chung}}, \bibinfo {author} {\bibfnamefont {M.~H.}\ \bibnamefont {Chen}}, \bibinfo {author} {\bibfnamefont {W.~L.}\ \bibnamefont {Morgan}}, \bibinfo {author} {\bibfnamefont {Y.}~\bibnamefont {Ralchenko}}, \ and\ \bibinfo {author} {\bibfnamefont {R.~W.}\ \bibnamefont {Lee}},\ }\bibfield  {title} {\enquote {\bibinfo {title} {{FLYCHK}: Generalized population kinetics and spectral model for rapid spectroscopic analysis for all elements},}\ }\href {\doibase 10.1016/j.hedp.2005.07.001} {\bibfield  {journal} {\bibinfo  {journal} {High Energy Density Physics}\ }\textbf {\bibinfo {volume} {1}},\ \bibinfo {pages} {3--12} (\bibinfo {year} {2005})}\BibitemShut {NoStop}%
\bibitem [{\citenamefont {Allais}\ \emph {et~al.}(2005)\citenamefont {Allais}, \citenamefont {Matte}, \citenamefont {Alouani-Bibi}, \citenamefont {Kim}, \citenamefont {Stotler},\ and\ \citenamefont {Rognlien}}]{Allais2005}%
  \BibitemOpen
  \bibfield  {author} {\bibinfo {author} {\bibfnamefont {F.}~\bibnamefont {Allais}}, \bibinfo {author} {\bibfnamefont {J.~P.}\ \bibnamefont {Matte}}, \bibinfo {author} {\bibfnamefont {F.}~\bibnamefont {Alouani-Bibi}}, \bibinfo {author} {\bibfnamefont {C.~G.}\ \bibnamefont {Kim}}, \bibinfo {author} {\bibfnamefont {D.~P.}\ \bibnamefont {Stotler}}, \ and\ \bibinfo {author} {\bibfnamefont {T.~D.}\ \bibnamefont {Rognlien}},\ }\bibfield  {title} {\enquote {\bibinfo {title} {Modification of atomic physics rates due to nonlocal electron parallel heat transport in divertor plasmas},}\ }\href {\doibase 10.1016/j.jnucmat.2004.10.089} {\bibfield  {journal} {\bibinfo  {journal} {Journal of Nuclear Materials}\ }\textbf {\bibinfo {volume} {337-339}},\ \bibinfo {pages} {246--250} (\bibinfo {year} {2005})}\BibitemShut {NoStop}%
\bibitem [{\citenamefont {Greenland}(2001)}]{Greenland2001}%
  \BibitemOpen
  \bibfield  {author} {\bibinfo {author} {\bibfnamefont {P.~T.}\ \bibnamefont {Greenland}},\ }\bibfield  {title} {\enquote {\bibinfo {title} {Collisional-radiative models with molecules},}\ }\href@noop {} {\ ,\ \bibinfo {pages} {1821--1839} (\bibinfo {year} {2001})}\BibitemShut {NoStop}%
\bibitem [{\citenamefont {Summers}\ \emph {et~al.}(2006)\citenamefont {Summers}, \citenamefont {Dickson}, \citenamefont {O'Mullane}, \citenamefont {Badnell}, \citenamefont {Whiteford}, \citenamefont {Brooks}, \citenamefont {Lang}, \citenamefont {Loch},\ and\ \citenamefont {Griffin}}]{Summers2006}%
  \BibitemOpen
  \bibfield  {author} {\bibinfo {author} {\bibfnamefont {H.~P.}\ \bibnamefont {Summers}}, \bibinfo {author} {\bibfnamefont {W.~J.}\ \bibnamefont {Dickson}}, \bibinfo {author} {\bibfnamefont {M.~G.}\ \bibnamefont {O'Mullane}}, \bibinfo {author} {\bibfnamefont {N.~R.}\ \bibnamefont {Badnell}}, \bibinfo {author} {\bibfnamefont {A.~D.}\ \bibnamefont {Whiteford}}, \bibinfo {author} {\bibfnamefont {D.~H.}\ \bibnamefont {Brooks}}, \bibinfo {author} {\bibfnamefont {J.}~\bibnamefont {Lang}}, \bibinfo {author} {\bibfnamefont {S.~D.}\ \bibnamefont {Loch}}, \ and\ \bibinfo {author} {\bibfnamefont {D.~C.}\ \bibnamefont {Griffin}},\ }\bibfield  {title} {\enquote {\bibinfo {title} {Ionization state, excited populations and emission of impurities in dynamic finite density plasmas: I. the generalized collisional-radiative model for light elements},}\ }\href {\doibase 10.1088/0741-3335/48/2/007} {\bibfield  {journal} {\bibinfo  {journal} {Plasma Physics and Controlled Fusion}\ }\textbf {\bibinfo {volume} {48}},\ \bibinfo {pages} {263--293} (\bibinfo {year} {2006})},\ \Eprint {http://arxiv.org/abs/astro-ph/0511561} {astro-ph/0511561} \BibitemShut {NoStop}%
\bibitem [{\citenamefont {Gu}(2004)}]{Gu2008}%
  \BibitemOpen
  \bibfield  {author} {\bibinfo {author} {\bibfnamefont {M.~F.}\ \bibnamefont {Gu}},\ }\bibfield  {title} {\enquote {\bibinfo {title} {The flexible atomic code},}\ }\href {\doibase 10.1139/P07-197} {\bibfield  {journal} {\bibinfo  {journal} {Canadian Journal of Physics}\ }\textbf {\bibinfo {volume} {86}},\ \bibinfo {pages} {675--689} (\bibinfo {year} {2004})},\ \bibinfo {note} {{ISBN}: 0735402116}\BibitemShut {NoStop}%
\bibitem [{\citenamefont {Lee}(1987)}]{Lee1987}%
  \BibitemOpen
  \bibfield  {author} {\bibinfo {author} {\bibfnamefont {Y.~T.}\ \bibnamefont {Lee}},\ }\bibfield  {title} {\enquote {\bibinfo {title} {A model for ionization balance and l-shell spectroscopy of non-{LTE} plasmas},}\ }\href {\doibase 10.1016/0022-4073(87)90039-2} {\bibfield  {journal} {\bibinfo  {journal} {Journal of Quantitative Spectroscopy and Radiative Transfer}\ }\textbf {\bibinfo {volume} {38}},\ \bibinfo {pages} {131--145} (\bibinfo {year} {1987})}\BibitemShut {NoStop}%
\bibitem [{\citenamefont {Marchand}, \citenamefont {Caillé},\ and\ \citenamefont {Lee}(1990)}]{Marchand1990}%
  \BibitemOpen
  \bibfield  {author} {\bibinfo {author} {\bibfnamefont {R.}~\bibnamefont {Marchand}}, \bibinfo {author} {\bibfnamefont {S.}~\bibnamefont {Caillé}}, \ and\ \bibinfo {author} {\bibfnamefont {Y.~T.}\ \bibnamefont {Lee}},\ }\bibfield  {title} {\enquote {\bibinfo {title} {Improved screening coefficients for the hydrogenic ion model},}\ }\href {\doibase 10.1016/0022-4073(90)90043-6} {\bibfield  {journal} {\bibinfo  {journal} {Journal of Quantitative Spectroscopy and Radiative Transfer}\ }\textbf {\bibinfo {volume} {43}},\ \bibinfo {pages} {149--154} (\bibinfo {year} {1990})}\BibitemShut {NoStop}%
\bibitem [{\citenamefont {Gaigalas}, \citenamefont {Rudzikas},\ and\ \citenamefont {Fischer}(1997)}]{Gaigalas1997}%
  \BibitemOpen
  \bibfield  {author} {\bibinfo {author} {\bibfnamefont {G.}~\bibnamefont {Gaigalas}}, \bibinfo {author} {\bibfnamefont {Z.}~\bibnamefont {Rudzikas}}, \ and\ \bibinfo {author} {\bibfnamefont {C.~F.}\ \bibnamefont {Fischer}},\ }\bibfield  {title} {\enquote {\bibinfo {title} {An efficient approach for spin-angular integrations in atomic structure calculations},}\ }\href {\doibase 10.1088/0953-4075/30/17/006} {\bibfield  {journal} {\bibinfo  {journal} {Journal of Physics B: Atomic, Molecular and Optical Physics}\ }\textbf {\bibinfo {volume} {30}},\ \bibinfo {pages} {3747--3771} (\bibinfo {year} {1997})}\BibitemShut {NoStop}%
\bibitem [{\citenamefont {Gaigalas}\ and\ \citenamefont {Fritzsche}(2002)}]{Gaigalas2002}%
  \BibitemOpen
  \bibfield  {author} {\bibinfo {author} {\bibfnamefont {G.}~\bibnamefont {Gaigalas}}\ and\ \bibinfo {author} {\bibfnamefont {S.}~\bibnamefont {Fritzsche}},\ }\bibfield  {title} {\enquote {\bibinfo {title} {Pure spin-angular momentum coefficients for non-scalar one-particle operators in jj-coupling},}\ }\href {\doibase 10.1016/S0010-4655(02)00589-1} {\bibfield  {journal} {\bibinfo  {journal} {Computer Physics Communications}\ }\textbf {\bibinfo {volume} {148}},\ \bibinfo {pages} {349--351} (\bibinfo {year} {2002})}\BibitemShut {NoStop}%
\bibitem [{\citenamefont {Bar-Shalom}, \citenamefont {Klapisch},\ and\ \citenamefont {Oreg}(1988)}]{Bar-Shalom1988}%
  \BibitemOpen
  \bibfield  {author} {\bibinfo {author} {\bibfnamefont {A.}~\bibnamefont {Bar-Shalom}}, \bibinfo {author} {\bibfnamefont {M.}~\bibnamefont {Klapisch}}, \ and\ \bibinfo {author} {\bibfnamefont {J.}~\bibnamefont {Oreg}},\ }\bibfield  {title} {\enquote {\bibinfo {title} {Electron collision excitations in complex spectra of ionized heavy atoms},}\ }\href {\doibase 10.1103/PhysRevA.38.1773} {\bibfield  {journal} {\bibinfo  {journal} {Physical Review A}\ }\textbf {\bibinfo {volume} {38}},\ \bibinfo {pages} {1773--1784} (\bibinfo {year} {1988})}\BibitemShut {NoStop}%
\bibitem [{\citenamefont {Oxenius}(1988)}]{Oxenius1988}%
  \BibitemOpen
  \bibfield  {author} {\bibinfo {author} {\bibfnamefont {J.}~\bibnamefont {Oxenius}},\ }\href {\doibase 10.3367/ufnr.0154.198803l.0536} {\emph {\bibinfo {title} {Kinetic theory of particles and photons}}}\ (\bibinfo  {publisher} {Springer-Verlag},\ \bibinfo {year} {1988})\ \bibinfo {note} {publication Title: Theoretical Foundations of Non-{LTE} Plasma Spectroscopy {ISSN}: 0042-1294}\BibitemShut {NoStop}%
\bibitem [{\citenamefont {Burgess}\ and\ \citenamefont {Chidichimo}(1983)}]{BurgessChidichimo1983}%
  \BibitemOpen
  \bibfield  {author} {\bibinfo {author} {\bibfnamefont {A.}~\bibnamefont {Burgess}}\ and\ \bibinfo {author} {\bibfnamefont {C.}~\bibnamefont {Chidichimo}},\ }\bibfield  {title} {\enquote {\bibinfo {title} {Electron impact ionization of complex ions},}\ }\href@noop {} {\bibfield  {journal} {\bibinfo  {journal} {Monthly Notices of the Royal Astronomical Society}\ }\textbf {\bibinfo {volume} {203}},\ \bibinfo {pages} {1269--1280} (\bibinfo {year} {1983})}\BibitemShut {NoStop}%
\bibitem [{\citenamefont {Regemorter}(1962)}]{Regemorter1962}%
  \BibitemOpen
  \bibfield  {author} {\bibinfo {author} {\bibfnamefont {H.~V.}\ \bibnamefont {Regemorter}},\ }\bibfield  {title} {\enquote {\bibinfo {title} {Rate of collisional excitation in stellar atmospheres},}\ }\href@noop {} {\bibfield  {journal} {\bibinfo  {journal} {The Astrophysical Journal}\ ,\ \bibinfo {pages} {906--915}} (\bibinfo {year} {1962})}\BibitemShut {NoStop}%
\bibitem [{\citenamefont {Kramers}(1923)}]{Kramers1923}%
  \BibitemOpen
  \bibfield  {author} {\bibinfo {author} {\bibfnamefont {H.~A.}\ \bibnamefont {Kramers}},\ }\href {\doibase 10.1080/14786442308565244} {\emph {\bibinfo {title} {{XCIII}. On the theory of X-ray absorption and of the continuous X-ray spectrum}}},\ Vol.~\bibinfo {volume} {46}\ (\bibinfo {year} {1923})\ \bibinfo {note} {publication Title: The London, Edinburgh, and Dublin Philosophical Magazine and Journal of Science Issue: 275 {ISSN}: 1941-5982}\BibitemShut {NoStop}%
\bibitem [{\citenamefont {Bates}, \citenamefont {Kingston},\ and\ \citenamefont {{McWhirter}}(1962)}]{Bates1962}%
  \BibitemOpen
  \bibfield  {author} {\bibinfo {author} {\bibfnamefont {D.~R.}\ \bibnamefont {Bates}}, \bibinfo {author} {\bibfnamefont {A.~E.}\ \bibnamefont {Kingston}}, \ and\ \bibinfo {author} {\bibfnamefont {W.~P.}\ \bibnamefont {{McWhirter}}},\ }\bibfield  {title} {\enquote {\bibinfo {title} {Recombination between electrons and atomic ions, 1. optically thin plasmas},}\ }\href@noop {} {\bibfield  {journal} {\bibinfo  {journal} {Proceedings of the Royal Society}\ }\textbf {\bibinfo {volume} {267}} (\bibinfo {year} {1962})}\BibitemShut {NoStop}%
\bibitem [{\citenamefont {Greenland}\ and\ \citenamefont {Reiter}(1998)}]{Greenland1998}%
  \BibitemOpen
  \bibfield  {author} {\bibinfo {author} {\bibfnamefont {P.~T.}\ \bibnamefont {Greenland}}\ and\ \bibinfo {author} {\bibfnamefont {D.}~\bibnamefont {Reiter}},\ }\bibfield  {title} {\enquote {\bibinfo {title} {Collisional radiative models with loss and recycling},}\ }\href {\doibase 10.1063/1.367513} {\bibfield  {journal} {\bibinfo  {journal} {Journal of Applied Physics}\ }\textbf {\bibinfo {volume} {83}},\ \bibinfo {pages} {7496--7503} (\bibinfo {year} {1998})}\BibitemShut {NoStop}%
\bibitem [{\citenamefont {Mijin}\ \emph {et~al.}(2021)\citenamefont {Mijin}, \citenamefont {Antony}, \citenamefont {Militello},\ and\ \citenamefont {Kingham}}]{Mijin2021}%
  \BibitemOpen
  \bibfield  {author} {\bibinfo {author} {\bibfnamefont {S.}~\bibnamefont {Mijin}}, \bibinfo {author} {\bibfnamefont {A.}~\bibnamefont {Antony}}, \bibinfo {author} {\bibfnamefont {F.}~\bibnamefont {Militello}}, \ and\ \bibinfo {author} {\bibfnamefont {R.~J.}\ \bibnamefont {Kingham}},\ }\bibfield  {title} {\enquote {\bibinfo {title} {{SOL}-{KiT}—fully implicit code for kinetic simulation of parallel electron transport in the tokamak scrape-off layer},}\ }\href {\doibase 10.1016/j.cpc.2020.107600} {\bibfield  {journal} {\bibinfo  {journal} {Computer Physics Communications}\ }\textbf {\bibinfo {volume} {258}},\ \bibinfo {pages} {107600} (\bibinfo {year} {2021})},\ \bibinfo {note} {publisher: Elsevier B.V.}\BibitemShut {Stop}%
\bibitem [{\citenamefont {Power}\ \emph {et~al.}(2021)\citenamefont {Power}, \citenamefont {Mijin}, \citenamefont {Militello},\ and\ \citenamefont {Kingham}}]{Power2021}%
  \BibitemOpen
  \bibfield  {author} {\bibinfo {author} {\bibfnamefont {D.}~\bibnamefont {Power}}, \bibinfo {author} {\bibfnamefont {S.}~\bibnamefont {Mijin}}, \bibinfo {author} {\bibfnamefont {F.}~\bibnamefont {Militello}}, \ and\ \bibinfo {author} {\bibfnamefont {R.~J.}\ \bibnamefont {Kingham}},\ }\bibfield  {title} {\enquote {\bibinfo {title} {Ion-electron energy transfer in kinetic and fluid modelling of the tokamak scrape-off layer},}\ }\href {\doibase 10.1140/epjp/s13360-021-02060-0} {\bibfield  {journal} {\bibinfo  {journal} {The European Physical Journal Plus}\ }\textbf {\bibinfo {volume} {136}},\ \bibinfo {pages} {1--13} (\bibinfo {year} {2021})},\ \bibinfo {note} {publisher: Springer Berlin Heidelberg {ISBN}: 1336002102060}\BibitemShut {NoStop}%
\bibitem [{\citenamefont {Braginskii}(1965)}]{Braginskii2004}%
  \BibitemOpen
  \bibfield  {author} {\bibinfo {author} {\bibfnamefont {S.~I.}\ \bibnamefont {Braginskii}},\ }\href {\doibase 10.1007/978-1-4757-4030-1_22} {\emph {\bibinfo {title} {Transport Processes in Plasmas}}}\ (\bibinfo {year} {1965})\ \bibinfo {note} {publication Title: Reviews of Plasma Physics}\BibitemShut {NoStop}%
\bibitem [{\citenamefont {Spitzer}\ and\ \citenamefont {Härm}(1953)}]{Spitzer1953}%
  \BibitemOpen
  \bibfield  {author} {\bibinfo {author} {\bibfnamefont {L.}~\bibnamefont {Spitzer}}\ and\ \bibinfo {author} {\bibfnamefont {R.}~\bibnamefont {Härm}},\ }\bibfield  {title} {\enquote {\bibinfo {title} {Transport phenomena in a completely ionized gas},}\ }\href {\doibase 10.1103/PhysRev.89.977} {\bibfield  {journal} {\bibinfo  {journal} {Physical Review}\ }\textbf {\bibinfo {volume} {89}},\ \bibinfo {pages} {977--981} (\bibinfo {year} {1953})}\BibitemShut {NoStop}%
\bibitem [{\citenamefont {Shkarofsky}, \citenamefont {Bachynski},\ and\ \citenamefont {Johnston}(1966)}]{shkarofsky1966}%
  \BibitemOpen
  \bibfield  {author} {\bibinfo {author} {\bibfnamefont {I.~P.}\ \bibnamefont {Shkarofsky}}, \bibinfo {author} {\bibfnamefont {M.~P.}\ \bibnamefont {Bachynski}}, \ and\ \bibinfo {author} {\bibfnamefont {T.~W.}\ \bibnamefont {Johnston}},\ }\href {https://books.google.co.uk/books?id=KZJTMwEACAAJ} {\emph {\bibinfo {title} {The Particle Kinetics of Plasmas}}}\ (\bibinfo  {publisher} {Reading, Mass.; Dordrecht printed},\ \bibinfo {year} {1966})\BibitemShut {NoStop}%
\bibitem [{\citenamefont {Stangeby}(2001)}]{Stangeby2001}%
  \BibitemOpen
  \bibfield  {author} {\bibinfo {author} {\bibfnamefont {P.~C.}\ \bibnamefont {Stangeby}},\ }\bibfield  {title} {\enquote {\bibinfo {title} {The plasma boundary of magnetic fusion devices},}\ }\href {\doibase 10.1088/0741-3335/43/2/702} {\bibfield  {journal} {\bibinfo  {journal} {Plasma Physics and Controlled Fusion}\ }\textbf {\bibinfo {volume} {43}},\ \bibinfo {pages} {223--224} (\bibinfo {year} {2001})}\BibitemShut {NoStop}%
\bibitem [{\citenamefont {Zhao}, \citenamefont {Chankin},\ and\ \citenamefont {Coster}(2019)}]{Zhao2019}%
  \BibitemOpen
  \bibfield  {author} {\bibinfo {author} {\bibfnamefont {M.}~\bibnamefont {Zhao}}, \bibinfo {author} {\bibfnamefont {A.~V.}\ \bibnamefont {Chankin}}, \ and\ \bibinfo {author} {\bibfnamefont {D.~P.}\ \bibnamefont {Coster}},\ }\bibfield  {title} {\enquote {\bibinfo {title} {Implementation of an inelastic collision operator into {KIPP}-{SOLPS} coupling and its effects on electron parallel transport in the scrape-off layer plasmas},}\ }\href {\doibase 10.1002/ctpp.201800130} {\bibfield  {journal} {\bibinfo  {journal} {Contributions to Plasma Physics}\ }\textbf {\bibinfo {volume} {59}},\ \bibinfo {pages} {1--13} (\bibinfo {year} {2019})}\BibitemShut {NoStop}%
\bibitem [{\citenamefont {Tskhakaya}(3 07)}]{tskhakaya_implementation_2023}%
  \BibitemOpen
  \bibfield  {author} {\bibinfo {author} {\bibfnamefont {D.}~\bibnamefont {Tskhakaya}},\ }\bibfield  {title} {\enquote {\bibinfo {title} {Implementation of dressed cross-section model into the {BIT}1 code},}\ }\href {\doibase 10.1140/epjd/s10053-023-00682-w} {\bibfield  {journal} {\bibinfo  {journal} {The European Physical Journal D}\ }\textbf {\bibinfo {volume} {77}},\ \bibinfo {pages} {135} (\bibinfo {year} {2023-07})}\BibitemShut {NoStop}%
\bibitem [{\citenamefont {Jayakumar}\ and\ \citenamefont {Fleischmann}(1985)}]{Jayakumar1985}%
  \BibitemOpen
  \bibfield  {author} {\bibinfo {author} {\bibfnamefont {R.}~\bibnamefont {Jayakumar}}\ and\ \bibinfo {author} {\bibfnamefont {H.~H.}\ \bibnamefont {Fleischmann}},\ }\bibfield  {title} {\enquote {\bibinfo {title} {Changes in spectral intensities of thermal plasmas in the presence of fast charged particles},}\ }\href {\doibase 10.1016/0022-4073(85)90103-7} {\bibfield  {journal} {\bibinfo  {journal} {Journal of Quantitative Spectroscopy and Radiative Transfer}\ }\textbf {\bibinfo {volume} {33}},\ \bibinfo {pages} {177--191} (\bibinfo {year} {1985})}\BibitemShut {NoStop}%
\bibitem [{\citenamefont {Mijin}\ \emph {et~al.}(2019)\citenamefont {Mijin}, \citenamefont {Militello}, \citenamefont {Newton}, \citenamefont {Omotani},\ and\ \citenamefont {Kingham}}]{Mijin2019}%
  \BibitemOpen
  \bibfield  {author} {\bibinfo {author} {\bibfnamefont {S.}~\bibnamefont {Mijin}}, \bibinfo {author} {\bibfnamefont {F.}~\bibnamefont {Militello}}, \bibinfo {author} {\bibfnamefont {S.}~\bibnamefont {Newton}}, \bibinfo {author} {\bibfnamefont {J.}~\bibnamefont {Omotani}}, \ and\ \bibinfo {author} {\bibfnamefont {R.~J.}\ \bibnamefont {Kingham}},\ }\href@noop {} {\enquote {\bibinfo {title} {Kinetic effects in parallel electron energy transport channels in the scrape-off layer},}\ } (\bibinfo {year} {2019})\BibitemShut {NoStop}%
\bibitem [{\citenamefont {Verhaegh}\ \emph {et~al.}(2023{\natexlab{a}})\citenamefont {Verhaegh}, \citenamefont {Lipschultz}, \citenamefont {Harrison}, \citenamefont {Osborne}, \citenamefont {Williams}, \citenamefont {Ryan}, \citenamefont {Allcock}, \citenamefont {Clark}, \citenamefont {Federici}, \citenamefont {Kool}, \citenamefont {Wijkamp}, \citenamefont {Fil}, \citenamefont {Moulton}, \citenamefont {Myatra}, \citenamefont {Thornton}, \citenamefont {Bosman}, \citenamefont {Bowman}, \citenamefont {Cunningham}, \citenamefont {Duval}, \citenamefont {Henderson},\ and\ \citenamefont {Scannell}}]{Verhaegh2023a}%
  \BibitemOpen
  \bibfield  {author} {\bibinfo {author} {\bibfnamefont {K.}~\bibnamefont {Verhaegh}}, \bibinfo {author} {\bibfnamefont {B.}~\bibnamefont {Lipschultz}}, \bibinfo {author} {\bibfnamefont {J.~R.}\ \bibnamefont {Harrison}}, \bibinfo {author} {\bibfnamefont {N.}~\bibnamefont {Osborne}}, \bibinfo {author} {\bibfnamefont {A.~C.}\ \bibnamefont {Williams}}, \bibinfo {author} {\bibfnamefont {P.}~\bibnamefont {Ryan}}, \bibinfo {author} {\bibfnamefont {J.}~\bibnamefont {Allcock}}, \bibinfo {author} {\bibfnamefont {J.~G.}\ \bibnamefont {Clark}}, \bibinfo {author} {\bibfnamefont {F.}~\bibnamefont {Federici}}, \bibinfo {author} {\bibfnamefont {B.}~\bibnamefont {Kool}}, \bibinfo {author} {\bibfnamefont {T.}~\bibnamefont {Wijkamp}}, \bibinfo {author} {\bibfnamefont {A.}~\bibnamefont {Fil}}, \bibinfo {author} {\bibfnamefont {D.}~\bibnamefont {Moulton}}, \bibinfo {author} {\bibfnamefont {O.}~\bibnamefont {Myatra}}, \bibinfo {author} {\bibfnamefont {A.}~\bibnamefont {Thornton}}, \bibinfo {author} {\bibfnamefont {T.~O. S.~J.}\ \bibnamefont {Bosman}}, \bibinfo {author} {\bibfnamefont {C.}~\bibnamefont {Bowman}}, \bibinfo {author} {\bibfnamefont {G.}~\bibnamefont {Cunningham}}, \bibinfo {author} {\bibfnamefont {B.~P.}\ \bibnamefont {Duval}}, \bibinfo {author} {\bibfnamefont {S.}~\bibnamefont {Henderson}}, \ and\ \bibinfo {author} {\bibfnamefont {R.}~\bibnamefont {Scannell}},\ }\bibfield  {title} {\enquote {\bibinfo {title} {Spectroscopic investigations of detachment on the {MAST} upgrade super-x divertor},}\ }\href@noop {} {\  (\bibinfo {year} {2023}{\natexlab{a}})},\ \bibinfo {note} {publisher: {IOP} Publishing}\BibitemShut {NoStop}%
\bibitem [{\citenamefont {Kobussen}(2023)}]{Kobussen2023}%
  \BibitemOpen
  \bibfield  {author} {\bibinfo {author} {\bibfnamefont {S.}~\bibnamefont {Kobussen}},\ }\bibfield  {title} {\enquote {\bibinfo {title} {Collisional radiative modelling with improved cross sections to investigate plasma molecular interactions in divertor plasmas},}\ }\href {http://arxiv.org/abs/2311.16732} {\  (\bibinfo {year} {2023})},\ \Eprint {http://arxiv.org/abs/2311.16732} {2311.16732} \BibitemShut {NoStop}%
\bibitem [{\citenamefont {Scarlett}\ \emph {et~al.}(2021)\citenamefont {Scarlett}, \citenamefont {Fursa}, \citenamefont {Zammit}, \citenamefont {Bray},\ and\ \citenamefont {Ralchenko}}]{Scarlett2021}%
  \BibitemOpen
  \bibfield  {author} {\bibinfo {author} {\bibfnamefont {L.~H.}\ \bibnamefont {Scarlett}}, \bibinfo {author} {\bibfnamefont {D.~V.}\ \bibnamefont {Fursa}}, \bibinfo {author} {\bibfnamefont {M.~C.}\ \bibnamefont {Zammit}}, \bibinfo {author} {\bibfnamefont {I.}~\bibnamefont {Bray}}, \ and\ \bibinfo {author} {\bibfnamefont {Y.}~\bibnamefont {Ralchenko}},\ }\bibfield  {title} {\enquote {\bibinfo {title} {Complete collision data set for electrons scattering on molecular hydrogen and its isotopologues},}\ }\href {\doibase 10.1016/j.adt.2020.101403} {\bibfield  {journal} {\bibinfo  {journal} {Atomic Data and Nuclear Data Tables}\ }\textbf {\bibinfo {volume} {139}},\ \bibinfo {pages} {101403} (\bibinfo {year} {2021})},\ \bibinfo {note} {publisher: Elsevier Inc.}\BibitemShut {Stop}%
\bibitem [{\citenamefont {Verhaegh}\ \emph {et~al.}(2023{\natexlab{b}})\citenamefont {Verhaegh}, \citenamefont {Lipschultz}, \citenamefont {Harrison}, \citenamefont {Federici}, \citenamefont {Moulton}, \citenamefont {Lonigro}, \citenamefont {Kobussen}, \citenamefont {O’Mullane}, \citenamefont {Osborne}, \citenamefont {Ryan}, \citenamefont {Wijkamp}, \citenamefont {Kool}, \citenamefont {Rose}, \citenamefont {Theiler},\ and\ \citenamefont {Thornton}}]{Verhaegh2023}%
  \BibitemOpen
  \bibfield  {author} {\bibinfo {author} {\bibfnamefont {K.}~\bibnamefont {Verhaegh}}, \bibinfo {author} {\bibfnamefont {B.}~\bibnamefont {Lipschultz}}, \bibinfo {author} {\bibfnamefont {J.}~\bibnamefont {Harrison}}, \bibinfo {author} {\bibfnamefont {F.}~\bibnamefont {Federici}}, \bibinfo {author} {\bibfnamefont {D.}~\bibnamefont {Moulton}}, \bibinfo {author} {\bibfnamefont {N.}~\bibnamefont {Lonigro}}, \bibinfo {author} {\bibfnamefont {S.}~\bibnamefont {Kobussen}}, \bibinfo {author} {\bibfnamefont {M.}~\bibnamefont {O’Mullane}}, \bibinfo {author} {\bibfnamefont {N.}~\bibnamefont {Osborne}}, \bibinfo {author} {\bibfnamefont {P.}~\bibnamefont {Ryan}}, \bibinfo {author} {\bibfnamefont {T.}~\bibnamefont {Wijkamp}}, \bibinfo {author} {\bibfnamefont {B.}~\bibnamefont {Kool}}, \bibinfo {author} {\bibfnamefont {E.}~\bibnamefont {Rose}}, \bibinfo {author} {\bibfnamefont {C.}~\bibnamefont {Theiler}}, \ and\ \bibinfo {author} {\bibfnamefont {A.}~\bibnamefont {Thornton}},\ }\bibfield  {title} {\enquote {\bibinfo {title} {The role of plasma-atom and molecule interactions on power \& particle balance during detachment on the {MAST} upgrade super-x divertor},}\ }\href {\doibase 10.1088/1741-4326/acf946} {\bibfield  {journal} {\bibinfo  {journal} {Nuclear Fusion}\ }\textbf {\bibinfo {volume} {63}} (\bibinfo {year} {2023}{\natexlab{b}}),\ 10.1088/1741-4326/acf946}\BibitemShut {NoStop}%
\bibitem [{\citenamefont {Tskhakaya}\ \emph {et~al.}(2011)\citenamefont {Tskhakaya}, \citenamefont {Jachmich}, \citenamefont {Eich},\ and\ \citenamefont {Fundamenski}}]{Tskhakaya2011}%
  \BibitemOpen
  \bibfield  {author} {\bibinfo {author} {\bibfnamefont {D.}~\bibnamefont {Tskhakaya}}, \bibinfo {author} {\bibfnamefont {S.}~\bibnamefont {Jachmich}}, \bibinfo {author} {\bibfnamefont {T.}~\bibnamefont {Eich}}, \ and\ \bibinfo {author} {\bibfnamefont {W.}~\bibnamefont {Fundamenski}},\ }\bibfield  {title} {\enquote {\bibinfo {title} {Interpretation of divertor langmuir probe measurements during the {ELMs} at {JET}},}\ }\href {\doibase 10.1016/j.jnucmat.2010.10.090} {\bibfield  {journal} {\bibinfo  {journal} {Journal of Nuclear Materials}\ }\textbf {\bibinfo {volume} {415}},\ \bibinfo {pages} {S860--S864} (\bibinfo {year} {2011})},\ \bibinfo {note} {publisher: Elsevier B.V.}\BibitemShut {Stop}%
\bibitem [{\citenamefont {Tskhakaya}(2017)}]{Tskhakaya2017}%
  \BibitemOpen
  \bibfield  {author} {\bibinfo {author} {\bibfnamefont {D.}~\bibnamefont {Tskhakaya}},\ }\bibfield  {title} {\enquote {\bibinfo {title} {One-dimensional plasma sheath model in front of the divertor plates},}\ }\href {\doibase 10.1088/1361-6587/aa8486} {\bibfield  {journal} {\bibinfo  {journal} {Plasma Physics and Controlled Fusion}\ }\textbf {\bibinfo {volume} {59}},\ \bibinfo {pages} {114001} (\bibinfo {year} {2017})},\ \bibinfo {note} {publisher: {IOP} Publishing}\BibitemShut {NoStop}%
\end{thebibliography}%

\end{document}